\documentclass[10pt, twocolumn, twoside]{IEEEtran}

\usepackage{amsmath,amssymb,graphicx,dsfont,cite,color}
\usepackage{enumerate, bm, upgreek, url, arydshln}

\newtheorem{theorem}{Theorem}
\newtheorem{lemma}{Lemma}

\newtheorem{cor}{Corollary}
\newtheorem{defn}{Definition}
\newtheorem{remark}{Remark}
\newtheorem{example}{Example}
\newtheorem{algorithm}{Algorithm}
\newtheorem{assumption}{Assumption}

\DeclareMathOperator*{\argmax}{arg\,max}

\begin{document}
	
\title{A Decentralized Optimal Feedback Flow Control Approach for Transport Networks}
\author{Saeid Jafari and Ketan Savla
\thanks{The authors are with the Sonny Astani Department of Civil and Environmental Engineering at the University of Southern California, Los Angeles, CA, 90089, USA. Email: {\tt\small\{sjafari,ksavla\}@usc.edu}.}
}
\maketitle
\bibliographystyle{IEEEtran}
\begin{abstract}
Finite-time optimal feedback control for flow networks under information constraints is studied. By utilizing the framework of multi-parametric linear programming, it is demonstrated that when cost/constraints can be modeled or approximated by piecewise-affine functions, the optimal control has a closed-form state-feedback realization. The optimal feedback control law, however, has a centralized structure and requires instantaneous access to the state of the entire network that may lead to prohibitive communication requirements in large-scale complex networks.  We subsequently examine the design of a decentralized optimal feedback controller with a one-hop information structure, wherein the optimum outflow rate from each segment of the network depends only on the state of that segment and the state of the segments immediately downstream. The decentralization is based on the relaxation of constraints that depend on state variables that are unavailable according to the information structure. The resulting decentralized control scheme has a simple closed-form representation and is scalable to arbitrary large networks; moreover, we demonstrate that, with respect to certain meaningful performance indexes, the performance loss due to decentralization is zero; namely, the centralized optimal controller has a decentralized realization with a one-hop information structure and is obtained at no computational/communication cost.
\end{abstract}
\IEEEpeerreviewmaketitle


\section{Introduction}\label{SEC_Intro}

In infrastructure flow networks with wide geographical distributions, a fast, efficient, and reliable control is essential. Examples of such networks include: transportation, natural-gas, water, and crude oil networks. In optimal network flow control, the primary objective is to regulate the flow in a transport network while optimizing a certain performance index. 

In the study of fluid dynamics at macroscopic scale, the fluid is treated as a continuum and its  motion is described by the \emph{mass conservation law} stating that ``the rate of change of the mass of a fluid in a fixed region is equal to the difference between the rate of mass flow into and out of the region'' \cite{Munson12}. Let $\rho(x,t)$ and $v(x,t)$ respectively denote the mass density and the velocity vector of a fluid at time $t$, at position $x=[x_1, x_2, x_3]^\top$ in the three-dimensional space. With the continuum representation of the fluid, the law of mass conservation is expressed as \cite{Munson12}:
\begin{equation}\label{MCL}
\begin{split}
\frac{\partial \rho}{\partial t}+\text{div}(\rho v)=0,
\end{split}
\end{equation}
which is balancing the rate of change of the mass density $\rho$ and the divergence of the mass flow rate  $\rho v$, where the \emph{divergence} of a vector field $f=[f_1, f_2, f_3]^\top$ in Cartesian coordinates is defined as $\text{div}(f) \triangleq \partial{f_1}/\partial{x_1}+\partial f_2/\partial {x}_2+\partial{f}_3/\partial{x}_3$. In order to simplify the analysis, fluid motion is often considered in one dimension reducing equation (\ref{MCL}) to 
\begin{equation}\label{MCL2}
\begin{split}
\frac{\partial \rho}{\partial t}=-\frac{\partial u}{\partial{x}},
\end{split}
\end{equation}
where $u=\rho{v}$ is the mass flow rate of the fluid. Many real flows are essentially one-dimensional, and variations in parameters across streamlines can be ignored; or by averaging properties of the flow over an appropriate region, it can be analyzed in one dimension \cite{Munson12}. In general, however, there are situations for which the one-dimensional assumption leads to highly erroneous results \cite{Munson12}.

\begin{figure}[h!]
\centering
\includegraphics[scale=0.26]{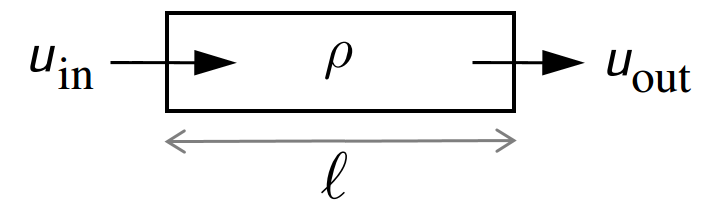}
\caption{One-dimensional fluid motion in a region (cell) of length $\ell$ and internal (average) mass density $\rho$ with inflow rate $u_{\text{in}}$ and outflow rate $u_{\text{out}}$.}
\label{fig1_gen}
\end{figure}

Now, consider fluid motion in a region (cell) of length $\ell$ as shown in Figure~\ref{fig1_gen} with inflow rate of $u_{\text{in}}$ into the cell and the outflow rate of $u_{\text{out}}$. 
A discretized version of (\ref{MCL2}), in both time and space, is given by
\begin{equation}\label{MCL3}
\begin{split}
\rho^{k+1}=\rho^k + \frac{T_s}{\ell}\left(u_{\text{in}}^k-u_{\text{out}}^k\right),
\end{split}
\end{equation}
where $T_s$ is the sampling time period, $\rho^k$ is the mass density of the fluid at time $t=kT_s$, and $u_{\text{in}}^k$  and $u_{\text{out}}^k$ are, respectively, the mass inflow and outflow rate into and from the cell at time step $t=kT_s$.

A widely-used approach for fluid flow control in a transport network is to partition the network into several segments, each of which is represented by a cell as shown in Figure~\ref{fig1_gen}. Then, the following assumptions are made: 
\begin{itemize}
	\item[(i)] The fluid dynamics in every cell is described by (\ref{MCL3}), that is, for cell $i$ of length $\ell_i$, mass density $\rho_i^k$, inflow rate $y_i^k$, and outflow rate $u_i^k$, we have
	\begin{equation}\label{CTM001}
	\begin{split}
	\rho_i^{k+1}=\rho_i^k + \frac{T_s}{\ell_i} (y_i^k-u_i^k).
	\end{split}
	\end{equation}
	\item[(ii)] The  mass density in every cell $\rho_i^k$ can be measured at each time step $k$.
	\item[(iii)] The outflow rate from each cell can be controlled through a regulation mechanism. This can be done by placing an \emph{active} network element (e.g. a control valve or a compressor) at interfaces between consecutive cells.
\end{itemize}

It should be noted that the inflow rate $y_i^k$ to cell $i$ is a known function of the outflow rates from the immediately upstream cells. If all immediate upstream cells of cell $i$ are merged only into cell $i$, then $y_i^k$ is equal to the sum of all flow rates leaving the upstream cells; otherwise, the inflow to cell $i$ is determined according to \emph{flow split ratios} of the network which are known \emph{a priori}. Hence, if the outflow rate from every cell is known over a fixed period of time, then from (\ref{CTM001}), the state of the system (densities) is completely known over that period.

\begin{figure}[h!]
	\centering
	\includegraphics[scale=0.35]{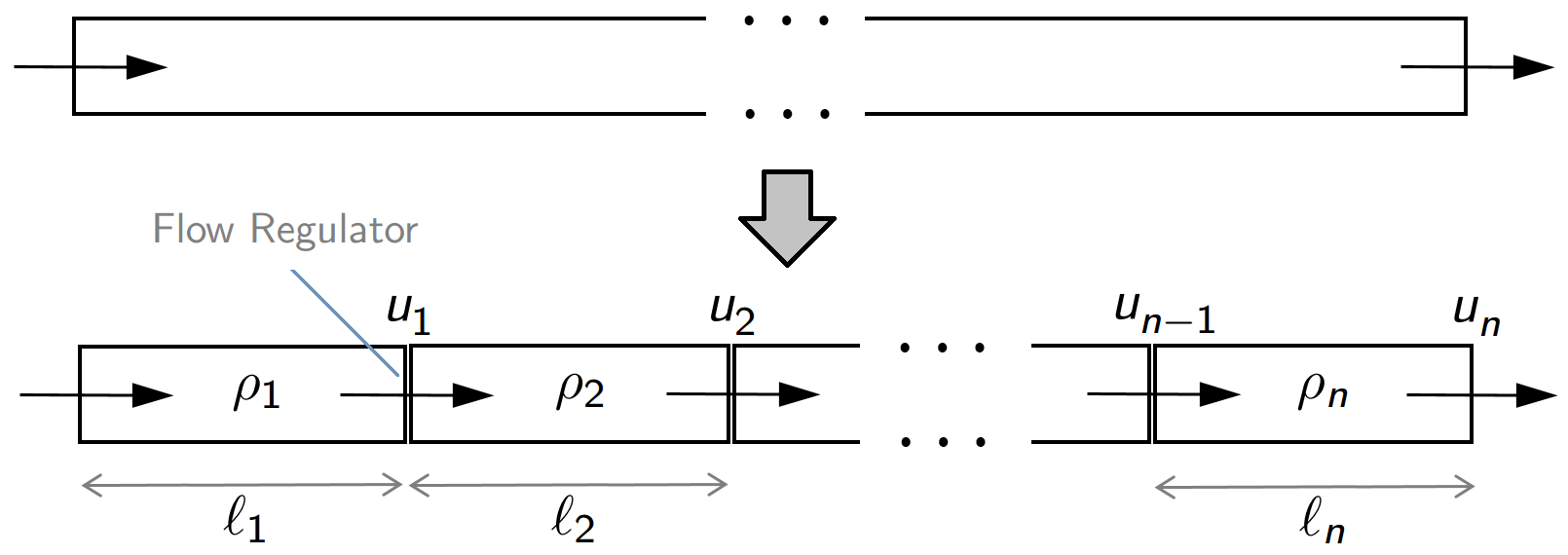}
	\caption{Partitioning a transmission network with a line structure into $n$ cells. A flow regulator at the interface between any two consecutive cells controls the outflow rate $u_i$ from each cell $i$. For this linear network, the inflow rate to cell $i$ is $y_i^k=u_{i-1}^k$.}
	\label{fig001}
\end{figure}

Figure~\ref{fig001} shows a fluid transmission network with a line structure partitioned into $n$ cells of possibly different length, where the cells are increasingly numbered from upstream to downstream. The outflow rate $u_i$ from cell $i$ can be controlled through a flow regulation mechanism. 

The control objective is to find time series of the outflow rates and the corresponding mass densities such as to optimize an integral performance index over a finite period of time, subject to dynamical and physical constraints of the network. In general, the optimization problem can be expressed as
\begin{equation}\label{genPerfIndex}
\begin{split}
&\min_{u} \left\{\varphi(\rho^N)+{\sum}_{k=0}^{N-1} \psi^k(\rho^k, u^k)\right\}\\
&\;\;\text{s.t.}\;(\rho, u)\in\Theta
\end{split},
\end{equation}
where $\rho=[\rho^{0^\top}, \ldots, \rho^{N^\top}]^\top$, $u=[u^{0^\top}, \ldots, u^{{(N-1)}^\top}]^\top$, $\rho^k=[\rho_1^k, \ldots, \rho_n^k]^\top$, $u^k=[u_1^k, \ldots, u_n^k]^\top$, $n$ is the number of cells, $N$ is the final time step, $\varphi$ is the terminal cost functional, $\psi$ is the running cost functional, and $\Theta$ is the set of admissible state/control variables satisfying (\ref{CTM001}) and meeting supply-demand constraints. The complexity of an optimization problem depends mainly on the function forms of its objective function and constraint set. For the sake of tractability, we focus on \emph{linear objective functions}. There are many meaningful cost functions of practical interest which can be expressed in a linear form \cite{Como16, Han2017, Muralidharan2015}.


The above framework has been widely used to formulate optimal flow control problems for complex transmission networks \cite{Daganzo94, Adacher2018, Wong1968, Misra15, Martin2006}. Two interesting examples are (i) highway traffic networks and (ii) natural gas transport networks. 

Traffic flow in highway transportation networks is often regulated by ramp metering and/or variable speed limit under the \emph{Cell Transmission Model} (CTM) dynamics. The CTM is a simple macroscopic traffic model capturing most phenomena observed on highways including flow conservation, non-negativity, and congestion wave propagation \cite{Daganzo94, Adacher2018}. Because of its analytical simplicity, the CTM is widely used for control design purposes, wherein a one-way road is partitioned into multiple cells as shown in Figure~\ref{fig001}, and the traffic flow in each cell is viewed as a homogeneous stream of vehicles with a dynamic described by (\ref{CTM001}). In this problem, the flow regulation is carried out by reducing the outflow rate from the cells, that is the flow regulation mechanism acts as a \emph{control valve}.

In a controlled natural gas transmission network with horizontal pipes, a pipeline consists of multiple champers (cells) with controllable outflow rates \cite{Wong1968}. The gas network's dynamic is  represented by (\ref{CTM001}) and flow regulation is performed through \emph{control valves} and \emph{compressors} \cite{Koch2015} to decrease and increase the flow rate, respectively such that desired supply and demand constraints are satisfied while a performance index of the form (\ref{genPerfIndex}) is optimized.  

Since the size and complexity of transportation networks are growing, design and implementation of an \emph{efficient} control scheme providing an optimum operation has become more challenging and demanding. The existing results on finite-time optimal control of transport networks are mainly restricted to schemes with an open-loop feedforward control structure which are not robust in most actual applications. It is well known that the use feedback helps reducing the effects of modeling uncertainties and improving performance, especially when a simplified plant model is used to make the control design and analysis tractable. 

One approach for optimal flow control is the \emph{Model Predictive Control} (MPC) which is a model-based feedback control technique relying on real-time optimization  \cite{Hegyi2005, Papamichail2010, Hadiuzzaman2013, Muralidharan2015, Han2017}. Although the closed-loop operation of the MPC provides a certain degree of robustness with respect to modeling uncertainties, the primary challenge of implementing MPC in real-time is its computational complexity. Moreover, determination of the optimal control action at each time step involves centralized operations making its implementation costly or impractical for large-scale networks.

It is, therefore, desired to design an optimal, or at least suboptimal, feedback control law with a simple structure that requires access only to local information. Decentralized optimal control problems are often substantially more complex than the corresponding problems with centralized information. A trivial centralized optimal decision-making problem may become NP-hard under a decentralized information structure \cite{Tsitsiklis_Athans_1985}. This is why most research has been focused on the design of meaningful suboptimal decentralized control policies and identification of tractable subclasses of problems \cite{Cogill2006, Lakshmanan2006}. Since no principled methodology exists for design and performance evaluation of decentralized optimal controllers, the problem is typically attacked by applying  suitable approximations and/or relaxations. 

This work is an attempt to deal with decentralized feedback control design for some classes of flow networks. A new decentralization method is proposed for feedback flow control, which is based on the following logic: (i) Construct a centralized optimal state-feedback control scheme with respect to a global performance index generating the control input of the entire network at each sample time, given the state vector of the entire network. The resulting controller, in theory, provides the ideal performance. In practice, however, such a controller may not be implementable. (ii) Design a local version of the centralized optimal feedback control scheme for each portion of the network minimizing a local cost function. The performance metric associated with each local controller is a local version of the global (centralized) performance index, wherein only local state variables (specified by a given information structure) are used to generate the input command to the respective actuator. Due to the lack of analytical tools, performance evaluation of the decentralized scheme and comparison with centralized optimal control are done through numerical simulations.

The rest of the paper is organized as follows: Section~\ref{SEC_preliminary} presents some preliminaries and notations used throughout the paper. A general formulation for control design procedure (centralized and decentralized) is presented in Section~\ref{SEC_controlDesign}. In Sections~\ref{SEC_gas} and \ref{SEC_traffic} flow control for  traffic and gas networks are studied. Numerical simulations are given in Section~\ref{SEC_simulation} to show the control performance, and finally concluding remarks are summarized in Section~\ref{SEC_conclusion}.


\section{Preliminaries and Notations}\label{SEC_preliminary}

Throughout this paper, the set of integers $\{1,2, \ldots, n\}$ is
denoted by $\mathbb{N}_n$, and $\{(a_i)_{i\in\mathbb{N}_n}\}=\{a_1, a_2, \ldots, a_n\}$. A \emph{convex polyhedron} is the intersection of finitely many half-spaces, i.e., $\{x\in\mathbb{R}^n\,|\,Ax\leq b \}$, for a matrix $A\in\mathbb{R}^{m\times n}$ and a vector $b\in\mathbb{R}^m$. A real-valued function $f(x)$ on $D\subseteq \mathbb{R}^n$ is said to be \emph{increasing} (\emph{decreasing}) if it is increasing (decreasing) in \emph{every} coordinate.


\begin{theorem}\cite{Borrelli03}\label{thm0}
	Consider the following multi-parametric linear program
	\begin{equation}\label{mp_LP}
	\begin{split}
	&J^*(\theta)=\min_z c^\top z\\
	&\text{s.t. } Wz\leq G+S\theta,\;\;\theta\in\Omega_\theta\subseteq \mathbb{R}^m,
	\end{split}
	\end{equation}
	where $z\in\mathbb{R}^n$ is the decision variables vector and $\theta\in\mathbb{R}^m$ is a parameter vector, $\Omega_\theta$ is a closed polyhedral set, and $c, W, G, S$ are constant matrices. Let $\Omega_\theta^*$ denote the region of parameters $\theta$ such that (\ref{mp_LP}) is feasible. Then, there exists an optimizer $z^*(\theta): \Omega_\theta^*\rightarrow\mathbb{R}^n$ which is a continuous and piecewise affine function of $\theta$, that is
	\begin{equation}\label{mp_sol}
	\begin{split}
	z^* &=\textit{pwa}(\theta)\\
	&=L_i\theta+l_i,\;\;\text{if } \theta\in\mathcal{R}_i,\;\;i\in\mathbb{N}_p,
	\end{split}
	\end{equation}
	where sets $\mathcal{R}_i=\{\theta\in\Omega_\theta^*\,|\, \Pi_i\theta\leq \eta_i\}$ form a polyhedral partition of $\Omega_\theta^*$, $p$ is the number of polyhedral sets, $L_i, l_i, \Pi_i, \eta_i$ are constant matrices, and $\textit{pwa}(\cdot)$ is a generic symbol for piecewise affine functions on polyhedral sets. Moreover, the value function $J^*(\theta): \Omega_\theta^*\rightarrow\mathbb{R}$ is a continuous, convex, and piecewise affine function of $\theta$.
\end{theorem}

The Matlab-based Multi-Parametric Toolbox \cite{MPT3} together with YALMIP Toolbox \cite{YALMIP_site} can be used to solve multi-parametric linear programs and compute the matrices $L_i, l_i, \Pi_i, \eta_i$ in (\ref{mp_sol}).

 
\section{Feedback Control Design}\label{SEC_controlDesign}
Consider the optimal control design problem (\ref{CTM001})-(\ref{genPerfIndex}). The objective is to design an optimal control with feedback architecture to benefit from the feedback properties such that the resulting control law is \emph{suitable} for practical implementation. By `suitable', we mean a controller meeting limitations in communication and computational power.

\begin{figure}[h!]
	\centering
	\includegraphics[scale=0.36]{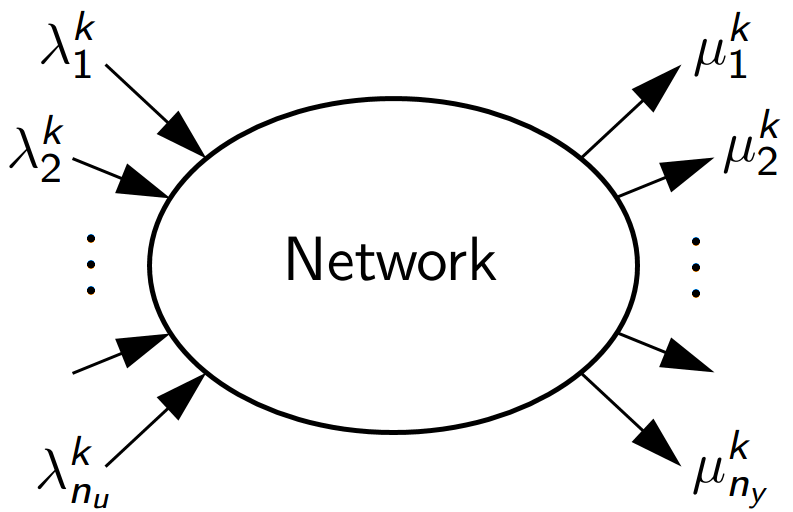}
	\caption{A general controlled network with $n_u$ exogenous inflow rates and $n_y$ outflow rates. The controller task is to regulate the outflow rate from each cell of the network to optimize a performance index.}
	\label{genNet1}
\end{figure}

Figure~\ref{genNet1} shows a general network with a number of inflow/outflow rates, where $\lambda_i^k$ and $\mu_i^k$ denote the $i$-th inflow and outflow at time $k$, respectively. The external inflow rates to the network act as exogenous inputs which cannot be manipulated by the controller. The controller can regulate only the outflow rate of each cell by monitoring the states of the network cells.  

Since the objective function and constraints in (\ref{genPerfIndex}) depend on inflow rates to the network, then, in general, a complete knowledge of inflow rate signals over the control horizon is required to solve the optimization problem (\ref{genPerfIndex}). The assumption that the external inflow rate over the control horizon is known \emph{a priori} is, however, very restrictive in practice. The exogenous input to the network may not be known or predictable in all scenarios. When no knowledge on the inflow rate is available, a control law must be designed such that the feasibility of the solution (control/state variables) at any time for \emph{any} admissible $\lambda_i^k$ is guaranteed. 

Disregarding communication and computational limitations, finding the global optimal to (\ref{genPerfIndex}) with a \emph{feedback realization} is a difficult task, if not impossible, in general. Hence, some assumptions and simplifications need to be made to make control design tractable. It is desired to implement the solution to (\ref{CTM001})-(\ref{genPerfIndex}) in the form of a \emph{static} state-feedback control as
\begin{equation}\label{centralizedGenFeedback}
\begin{split}
u^k=\phi^k(\rho^k)
\end{split},
\end{equation}
where $u^k=[u_1^k, \ldots, u_n^k]^\top$ and $\rho^k=[\rho_1^k, \ldots, \rho_n^k]^\top$ are the vectors of cells' outflow rates and mass densities, respectively. A realization of the form (\ref{centralizedGenFeedback}) is possible when the performance index and constraints satisfy certain properties, or they are simplified through proper approximations to satisfy certain properties.

\begin{remark}
	The main reason for considering ``static feedback'' is the simplicity of control law. In a static state-feedback controller, the control action at each time $k$ depends only on the current state vector at time $k$. One may consider a ``dynamic feedback'' controller, wherein the control action depends on the state variables in the current and previous sampling instants; this, however, makes design, analysis, and implementation of the controller more difficult. 
\end{remark}

Although design of a centralized feedback optimal control (if it exists) provides the ideal performance, it may not be implementable for large-size networks, as it may require a significant computational resource and a fast and highly-reliable commutation system. It is, therefore, necessary to further simplify the control law to meet communication/computational constraints.

In order to address the aforementioned problems and to examine the properties of controlled flow networks, we focus on two classes of networks: (i) \emph{highway traffic networks} modeled by the CTM and (ii) \emph{natural gas transmission networks}, both of which can be formulated as (\ref{CTM001})-(\ref{genPerfIndex}). 

We first study the centralized control design under certain assumptions such that the control law is optimal (in some sense) and implementable in a state-feedback form. Subsequently, decentralization of the resulting centralized control scheme is investigated by considering a simple information structure. Communication constraints are often modeled by a fixed \emph{information structure}. For example, in the network shown in Figure~\ref{fig001}, if to generate the outflow from every cell $i$, only the mass density of cells $i$ and $i+1$ are available for measurement, a desired decentralized realization of the $i$-th controller is
\begin{equation}\label{decentralizedGenFeedback}
\begin{split}
u_i^k=\phi_i^k(\rho_i^k, \rho_{i+1}^k).
\end{split}
\end{equation}

\begin{remark}
	It is to be noted that (\ref{centralizedGenFeedback}) and (\ref{decentralizedGenFeedback}) do not necessarily mean that the feedback controller must possess a closed-form expression. They can be viewed as algorithms whose input is (local) state vector at time $k$ and its output is the (local) control action at time $k$. 
\end{remark}

We argue that when the cost and constraint functions satisfy some separability condition, a local version of problem (\ref{CTM001})-(\ref{genPerfIndex}) can be constructed for each portion of the network. A local area of a controller/actuator is determined by a given information structure restricting the knowledge of each local controller on the network's state. That is, the $i$-th local controller (generating the outflow rate from cell $i$) has access only to the state of cells in a pre-specified neighborhood of cell $i$.  By expressing the solution to each local problem in a feedback form, a state-feedback decentralized control law is obtained. It should be highlighted that to design a local controller no information about the parameters and state of the rest of the network is available; this is the feedback architecture of the control law that indirectly provides some information about non-local cells. In other words,  \emph{feedback} is essential to keep a local controller from being completely blind about the rest of the network.

\begin{figure}[h!]
	\centering
	\includegraphics[scale=0.38]{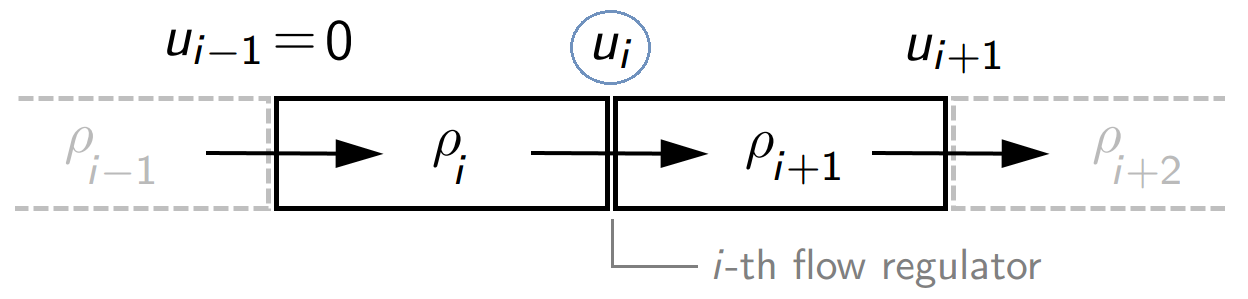}
	\caption{To design the $i$-th controller generating outflow rate $u_i^k$, a centralized optimal feedback control law for the sub-network consisting of only cells $i$ and $i+1$ is designed. No knowledge about the rest of the network is available to the $i$-th controller.}
	\label{twoCellNet}
\end{figure}

To further illustrate the proposed decentralization, let us consider the network shown in Figure~\ref{fig001}, and assume that only knowledge about cells $i$ and $i+1$ are available to generate $u_i$. To design the $i$-th controller, we consider the sub-network consisting of only cells $i$ and $i+1$, as shown in Figure~\ref{twoCellNet} and solve the centralized problem associated with the two-cell network. In the $i$-th local optimization problem, the decision variables are $u^k_i, u^k_{i+1}$, $k=0, \ldots, N-1$, with zero inflow rate to cell $i$, but only $u_i^k$ is used and implemented and the optimal values of $u^k_{i+1}$ are unused. For this example, the $i$-th local optimization problem may be expressed as
\begin{align}
&\min_{u_i, u_{i+1}} \Big{\{}{\hat\varphi}_i(\rho_i^N, \rho_{i+1}^N) + {\sum}_{k=0}^{N-1} \hat{\psi}_i^k(\rho_i^k, \rho_{i+1}^k, u_i^k, u_{i+1}^k)\Big{\}}\nonumber\\
&\;\;\;\;\;\;\text{s.t.}\;(u_i, u_{i+1}, \rho_i, \rho_{i+1})\in{\hat{\Theta}}_i\label{localPerfIndex}
\end{align}
where $\hat\varphi_i$ is the terminal cost functional, $\hat{\psi}_i$ is the running cost functional, and ${\hat{\Theta}}_i$ is the outflow constraint set associated with the $i$-th controller. The set ${\hat{\Theta}}_i$ is obtained by relaxing any constraint involving non-local variables. From the solutions to (\ref{localPerfIndex}), only $u_i^k$ is kept for implementation and the remaining variables ($u^k_{i+1}$) are discarded.  As mentioned before, we would like to implement the solution to (\ref{localPerfIndex}) in the form of a  static state-feedback of local states as (\ref{decentralizedGenFeedback}). The feedback realization of the solution is crucial as the values of  $\rho_i$ and $\rho_{i+1}$ are affected by the action of the other controllers in the network.

The proposed decentralization scheme relies on the following properties:
\begin{itemize}
	\item Existence of a global optimizer for the centralized problem with a state-feedback realization (not necessarily in a closed form).
	\item Separability of the centralized optimization problem such that a well-defined local version of the centralized optimal control problem can be constructed for each sub-network for which a global optimizer can be found in a feedback form.
\end{itemize}


\section{Traffic Network}\label{SEC_traffic}

In recent years, due to the ever-increasing traffic demand, efficient control and management of transportation networks has received a great deal of attention. There has been a lot of research done on the optimal control of freeway networks based on various models for traffic systems, among which first-order models, such as the CTM, are widely used for control design. In a CTM-based traffic model, the network dynamics  is described by (\ref{CTM001}), where $\rho_i^k$ [veh/mi] is the traffic density, $y_i^k$ [veh/hr] is the inflow rate, $u_i^k$ [veh/hr] is the outflow rate, and $\ell_i$ [mi] is the length of cell $i$. The constraints are defined in terms of demand and supply functions, where the demand function $\bar{d}_i(\cdot)$ returns the maximum outflow from the cell as a function of its current traffic density, and the supply function $\bar{s}_i(\cdot)$ gives the maximum inflow into the cell as a function of its current traffic density \cite{Como16}. The demand and supply functions are assumed to be of the form
\begin{equation}\label{const001}
\begin{split}
&\bar{d}_i(\rho_i) = \min\{d_i(\rho_i), C_i\},\\
&\bar{s}_i(\rho_i) = \min\{s_i(\rho_i), C_i\},
\end{split}
\end{equation}
where $d_i$ is continuous non-decreasing function of $\rho_i$ with $d_i(0)=0$ and $s_i$ is continuous non-increasing function of $\rho_i$ with $s_i(0)>0$, and $C_i$  [veh/hr] is maximum flow capacity of cell $i$. The \emph{jam traffic density} of cell $i$ is defined as $\gamma_i=\inf\{\rho_i>0\,|\,s_i(\rho_i)=0\}$. The functions $d_i(\cdot)$ and $s_i(\cdot)$ are often assumed to be affine of the form $d_i(\rho_i)=v_i \rho_i$ and $s_i(\rho_i)=w_{i}(\gamma_{i}-\rho_{i})$, where  $v_i$ [mi/hr] is the maximum traveling free-flow speed and $w_i$ [mi/hr] is the backward congestion wave traveling speed of cell $i$. Then, in a \emph{controlled network} via speed limit control, the feasible region for outflow and inflow rates are defined as \cite{Como16}:
\begin{equation}\label{const1}
\begin{split}
&0\leq u_i^k\leq \min\{v_i \rho_i^k, C_i\},\\
&0\leq y_i^k\leq \min\{w_{i}(\gamma_{i}-\rho_{i}^k), C_{i}\}.
\end{split}
\end{equation}

The flow regulation mechanism in a traffic network acts as a collection of  \emph{control valves}, each of which at each time can be open to the fullest extent possible, completely closed, or partially closed during the network operation. 


\begin{assumption}\label{assumption01}
	The length of cells $\ell_i$ and the time interval $T_s$ are chosen such that vehicles traveling at maximum speed $v_i$ can not cross multiple cells in one time step, i.e., $v_i T_s\leq \ell_i$, $\forall i$. Also, the backward congestion wave traveling speed $w_i$ satisfies $w_i T_s\leq \ell_i$, $\forall i$.
\end{assumption}

Assumption~\ref{assumption01} is known as \emph{Courant-Friedrichs-L\`{e}vy} condition \cite{Como16} which is a necessary condition for numerical stability in numerical computations. It can be easily verified that Assumption~\ref{assumption01} together with constraints (\ref{const1}) ensure that at each time the density of each cell is non-negative and never exceeds the jam density.

A flow network can be represented by a directed graph, in which edges represent cells and vertices (or junctions) represent interface between consecutive cells which are the actuators' location. The junctions can be of either of the three types defined below.
\begin{defn} \cite{Como16} \label{def01}
	A junction with a single incoming and a single outgoing cell is called \emph{ordinary}; a junction with a single incoming cell and multiple outgoing cells is called \emph{diverge}; and a junction with multiple incoming cells and a single outgoing cell is called \emph{merge}.
\end{defn}

The following definitions and notations are used throughout this section.
\begin{defn}
	Consider a  network whose topology is described by directed graph $\mathcal{G}$. The set of edges of $\mathcal{G}$ corresponding to on-ramps is called the \emph{source set} denoted by $\mathcal{E}_{\text{on}}$, and the set of edges corresponding to off-ramps is called the \emph{sink set} denoted by $\mathcal{E}_{\text{off}}$.
\end{defn}

At any diverge  junction, the traffic flow is distributed according to a given split percentage which are estimated from historical data \cite{Krumm10}.

\begin{defn} \cite{Como16} \label{turning_ratio_def}
	The \emph{split ratio} (or \emph{turning ratio}) $R_{ij}\in[0,1]$ is defined as the fraction of flow leaving cell $i\not\in\mathcal{E}_{\text{off}}$ that is directed towards cell $j\neq i$, where ${\textstyle\sum}_j R_{ij}=1$. If cells $i$ and $j$ are not adjacent or $i=j$, $R_{ij}$ is defined to be zero.
\end{defn}

\begin{defn} \label{outneighbor}
	Let cell $i$ be an incoming cell to junction $\hbar_i$, where $\hbar_i$ denotes the \emph{head} or the \emph{downstream junction} of cell $i$. The set of all outgoing cells from junction $\hbar_i$ is called the \emph{out-neighborhood} of cell $i$ and is denoted by $\mathcal{E}_i^+$. If $i\in\mathcal{E}_{\text{off}}$, then $\mathcal{E}_i^+$ is the empty set. In other words, $\mathcal{E}_i^+$ is the set of all direct successor of cell $i$. The elements of $\mathcal{E}_i^+$ are referred to as the \emph{out-neighbors} of cell $i$.
\end{defn}

\begin{defn} \label{inneighbor} \label{def05}
	Let cell $i$ be an outgoing cell from junction $\tau_i$,  where $\tau_i$ denotes the \emph{tail} or the \emph{upstream junction} of cell $i$. The set of all incoming cells to junction $\tau_i$ is called the \emph{in-neighborhood} of cell $i$ and is denoted by $\mathcal{E}_i^-$.  If $i\in\mathcal{E}_{\text{on}}$, then $\mathcal{E}_i^-$ is the empty set. In other words, $\mathcal{E}_i^-$ is the set of all direct predecessor of cell $i$. The elements of $\mathcal{E}_i^-$ are referred to as the \emph{in-neighbors} of cell $i$.
\end{defn}

\begin{figure}[h!]
	\centering
	\includegraphics[scale=0.41]{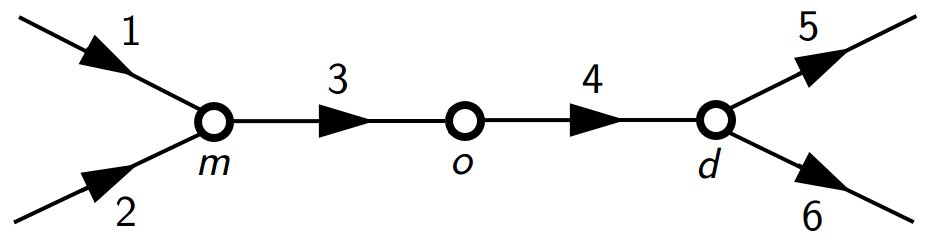}
	\caption{Directed graph of a six-cell network with source set (on-ramps) $\mathcal{E}_{\text{on}}=\{1,2\}$ and sink set (off-ramps) $\mathcal{E}_{\text{off}}=\{5,6\}$. The merge, ordinary, and diverge junctions are labeled $m$, $o$, and $d$, respectively. The split ratios of $R_{45}=0.3$ and $R_{46}=0.7$ imply that $30\%$ of vehicles in cell $4$ turn towards cell $5$ and $70\%$ of them turn toward cell $6$. The in-neighborhood and out-neighborhood of cell $4$ are   $\mathcal{E}_{4}^-=\{3\}$ and $\mathcal{E}_{4}^+=\{5,6\}$, respectively.}
	\label{fig_gen_net}
\end{figure}

An example is shown in Figure~\ref{fig_gen_net} clarifying the above definitions.

It is often more convenient to express the dynamics and constraints in terms of the \emph{traffic mass} of the cells. Let $x_i^k=\ell_i \rho_i^k$ [veh] denote the traffic mass of cell $i$ at time $k$, then from (\ref{CTM001}) and (\ref{const1}), the dynamics and constraints of an $n$-cell network can be written as
\begin{subequations}\label{Gen_net}
	\begin{align}
	&x^{k+1}_i=x^k_i+T_s(y^k_{i}-u^k_i),\;\;\;\forall i\in\mathbb{N}_n \label{Gen_net_a}\\
	&y_i^k=\lambda_i^k+ {\textstyle\sum}_{j=1}^n R_{ji}u_j^k, \label{Gen_net_b}\\
	&0\leq u_i^k\leq \min\{(v_i/\ell_i)x_i^k, C_i\},\\
	&0\leq y_i^k\leq \min\{w_{i}(\gamma_{i}-(1/\ell_{i})x_{i}^k), C_{i}\},
	\end{align}
\end{subequations}
where $\lambda_i^k$ is an exogenous inflow rate to cell $i\in\mathcal{E}_{\text{on}}$ ($\lambda_i^k=0$, if $i\not\in\mathcal{E}_{\text{on}}$), and $R_{ij}$'s are split ratios. Then, for any $i\in\mathcal{E}_{\text{on}}$, $y_i^k=\lambda_i^k$, and for any  $i\not\in\mathcal{E}_{\text{on}}$, $y_i^k={\textstyle\sum}_{j=1}^n R_{ji}u_j^k$.

\begin{remark}\label{onramp_assumption}
To ensure that $\lambda_i^k$ is a feasible exogenous input to the network, it is typically assumed that the jam traffic density of any on-ramp is infinity, $\gamma_i=\infty$, and $\lambda_i^k\leq C_i$, $\forall i\in\mathcal{E}_{\text{on}}$.   
\end{remark}

\emph{Control Objective}: Consider the network dynamics (\ref{Gen_net}) and let $x^k=[x_1^k, \ldots, x_n^k]^\top$ be the state vector and $u^k=[u_1^k, \ldots, u_n^k]^\top$ be the control input vector of the network at time $k$. The control objective is to design a static feedback control law such that for \emph{any} initial state $x^0$ and \emph{any} exogenous inflow $\lambda^k$, the feasibility of control actions is guaranteed and a performance index of the form (\ref{genPerfIndex}), subject to (\ref{Gen_net}) and a given information structure, over a fixed given control horizon $[0, N]$ is optimized. In this paper, we focus on linear  objective functions, i.e. (\ref{genPerfIndex}) with
\begin{equation}\label{general_linear_cost}
\begin{gathered}
\varphi(x^N)={\textstyle\sum}_{i=1}^n \alpha_i^Nx_i^N,\\
\psi^k(x^k,u^k)={\textstyle\sum}_{i=1}^n \alpha_i^k x_i^k + \beta_i^ku_i^k,
\end{gathered}
\end{equation}
where $N$ is a fixed final time, and $\alpha_i^k\geq 0$ and $\beta_i^k$ are cost-weighting parameters.

\begin{remark}\label{REM_special_costs}
	There are meaningful performance indexes which can be expressed in a linear form \cite{Como16, Han2017, Muralidharan2015}; for example:
	\begin{itemize}
		\item[(i)] Minimization of the \emph{total travel time} of the network is equivalent to minimization of the total number of vehicles in the entire network, then the corresponding cost is $J = {\textstyle\sum}_{k=0}^N{\textstyle\sum}_{i=1}^n x_i^k$.
		\item[(ii)] Maximization of the \emph{total travel distance} is equivalent to maximization of the flows, then the following cost should be minimized $J = -{\textstyle\sum}_{k=0}^{N-1}{\textstyle\sum}_{i=1}^n u_i^k$.
		\item[(iii)] The \emph{total congestion delay} is defined as the time difference between actual travel time and the travel time in free-flow conditions whose minimization is equivalent to minimizing $J = {\textstyle\sum}_{k=0}^{N-1}{\textstyle\sum}_{i=1}^n (x_i^k-(\ell_i/v_i)u_i^k)$.
	\end{itemize}
\end{remark}

For the centralized control, there is no information constraint and the control law is of the form $u^k=\phi^k(x^k)$. For the decentralized control, we consider controller with a \emph{one-hope information structure} as defined below.

\begin{defn} \label{one_hop_def}
	A feedback controller is said to have a \emph{one-hop information structure}, if $u_i^k$ depends only on $x_i^k$ and the state of the cell(s) immediately downstream of cell $i$, i.e. those either entering or leaving the downstream junction of cell $i$. 
\end{defn}

For the network in Figure~\ref{fig_gen_net}, a decentralized static feedback controller with a one-hop information structure is of the form: $u_1^k=\phi_1^k(x_1^k, x_2^k, x_3^k)$, $u_2^k=\phi_2^k(x_2^k, x_1^k, x_3^k)$, $u_3^k=\phi_3^k(x_3^k, x_4^k)$, $u_4^k=\phi_4^k(x_4^k, x_5^k, x_6^k)$, $u_5^k=\phi_5^k(x_5^k)$, and $u_6^k=\phi_6^k(x_6^k)$.

\subsection{Centralized Feedback Control}\label{SEC_centralized}

The external inflow rate $\lambda_i^k$ to the network acts as an exogenous input which cannot be manipulated by the controller. The solution to the optimization problem (\ref{genPerfIndex}), (\ref{Gen_net}), (\ref{general_linear_cost}) depends, in general, on the values of $\lambda_i^k$; however, no \emph{a priori} knowledge on $\lambda_i^k$ is often available for control design. Analogous to the standard LQR problem where no uncontrolled exogenous input is assumed for optimal control design, we design a centralized optimal controller under the assumption of $\lambda_i^k=0$, $\forall i\in\mathcal{E}_{\text{on}},  k\in[0, N-1]$; and we refer to the resulting controller as \emph{zero-$\lambda$ optimal control law}. Then, we show that the feasibility of the optimal solution in guaranteed for any nonzero $\lambda_i^k$.

Let us first suppose that the sequence of $\lambda_i^k$, $\forall i\in\mathcal{E}_{\text{on}}$ over the control horizon is known beforehand. The following theorem gives the \emph{true optimal control law}.

\begin{theorem}\label{thm1}
	The solution to (\ref{genPerfIndex}), (\ref{Gen_net}), (\ref{general_linear_cost}) can be expressed in the form of a continuous piecewise affine static feedback law on polyhedra of the state vector as
	\begin{equation}\label{centralized_controller}
	\begin{split}
	(u^k)^* &=\textit{pwa}^k(x^k)\\
	&={F}_i^kx^k + {f}^k_i,\;\;\text{if } x^k\in\mathcal{R}_i^k,
	\end{split}
	\end{equation}
	where $\mathcal{R}_i^k=\{x\in\mathbb{R}^n\,|\,{H}^k_ix\leq {h}^k_i\}$, $i\in\mathbb{N}_{p^k}$, is the $i$-th polyhedral partition of the set of feasible states, and $p^k$ is the number of polyhedral partitions at time $k$. The controller parameters ${F}^k_i, {f}^k_i, {H}^k_i, {h}^k_i$ can be computed offline; they are independent of $x^k$, $\forall k$, but may depend on the values of $\lambda_i^k$.
\end{theorem}

\emph{Proof}: The proof is given in the Appendix.

\begin{cor}\label{cor01}
	Consider the optimization problem (\ref{genPerfIndex}), (\ref{Gen_net}), (\ref{general_linear_cost}). The \emph{zero-$\lambda$ optimal control law} can be expressed as (\ref{centralized_controller}), where matrices ${F}^k_i, {f}^k_i, {H}^k_i, {h}^k_i$ can be computed off-line. Moreover, the feasibility of the resulting control actions is guaranteed for any non-zero inflow rate $\lambda_i^k$, i.e., the constraints in (\ref{Gen_net}) are always satisfied.
\end{cor}

\emph{Proof}: The proof is given in the Appendix.

With respect to certain cost functions, the zero-$\lambda$ optimal feedback control law is truly optimal. We show that for problem (\ref{genPerfIndex}), (\ref{Gen_net}), (\ref{general_linear_cost}), under certain assumptions on the network topology, if the cost functions satisfy certain properties, the centralized optimal feedback controller has a decentralized realization with a one-hop information structure which is truly optimal for any exogenous inflow $\lambda_i^k$.

\begin{theorem}\label{thm_GEN}
	Consider the problem (\ref{genPerfIndex}), (\ref{Gen_net}), (\ref{general_linear_cost}) for a network with time-invariant split ratios and \emph{no merge junction}. In addition, assume that cost-weighting parameters satisfy $\alpha_i^k\geq \alpha_{j}^k\geq 0$, $\forall k, i$ and $\forall j\in\mathcal{E}_i^+$, and $\beta_i^k\leq \beta_i^{k+1}\leq 0$, $\forall k, i$. Then, the true optimal feedback control law (with centralized information) can be realized as
	\begin{align}
	(u_i^k)^* &= \textit{pwa}^k_i(x_i^k, (x_{j}^k)_{j\in\mathcal{E}_i^+})\label{decentReal}\\
	&= \min\left\{\frac{v_i}{\ell_i}x_i^k, C_i,\Big{(}  \frac{w_{j}}{R_{ij}}(\gamma_{j}-\frac{1}{\ell_{j}}x_{j}^k), \frac{C_{j}}{R_{ij}}\Big{)}_{j\in\mathcal{E}_i^+}\right\}.\nonumber
	\end{align}
\end{theorem}

\emph{Proof}: The proof is given in the Appendix.

The controller (\ref{decentReal}) has a one-hop information structure (see Definition~\ref{one_hop_def}); moreover, its parameters are obtain at \emph{no computational cost} independent of the control horizon $N$. Indeed, the expression in the right-hand side of (\ref{decentReal}) is the upper limit of $u_i^k$ which is known beforehand, that is $0\leq u_i^k\leq \bar{u}_i^k$, where 
\begin{equation}
\bar{u}_i^k \!=\! \min\left\{\!\frac{v_i}{\ell_i}x_i^k, C_i,\Big{(}  \frac{w_{j}}{R_{ij}}(\gamma_{j}\!-\!\frac{1}{\ell_{j}}x_{j}^k), \frac{C_{j}}{R_{ij}}\Big{)}_{j\in\mathcal{E}_i^+}\!\right\}.
\end{equation}
Hence, Theorem~\ref{thm_GEN} states that, under the given assumptions, setting each outflow rate equal to its upper limit provides the true optimal performance. This is equivalent to opening every control valve to the fullest extent possible at each time. We refer to such scheme as  \emph{trivial control} or \emph{uncontrolled scheme}. 

\begin{remark}
	The conditions given in Theorem~\ref{thm_GEN} are sufficient (not necessary) on a linear performance index with respect to which a centralized optimal control law has a realization with a one-hop information structure. 
	Moreover, the optimal control is not necessarily unique.
\end{remark}

\begin{remark}
	The widely-used performance indexes in Remark~\ref{REM_special_costs} satisfy the properties given in Theorem~\ref{thm_GEN}. 
\end{remark}

In general, however, the optimal controller needs access to the state of the entire network and depends on the control horizon. For a general network with a general linear cost functional, the closed-form of the control law (\ref{centralized_controller}) enables one to compute the controller parameters offline and stored in computer memory before the control actions are ever applied to the network. That is, there is no need to solve a large-size optimization problem at every time step for real-time implementation, unless there is a large variation in the network parameters. The optimal feedback controller (\ref{centralized_controller}), however, suffers from two major drawbacks restricting its applicability to large-scale networks: (i) Even though the piecewise affine form of the control law seems to be simple, when the number of cells and the control horizon increase, solving the corresponding multi-parametric linear programs may result in a very large number of polyhedral partitions, making the structure of the controller too complex. Although applying the merging algorithms \cite{MPT3, Baotic2003} may considerably reduce the number of polyhedral partitions, in general there may still be too many polyhedral sets. (ii) Determining the optimal control action at each time involves centralized operations, that is each local controller needs instantaneous access to the state of the entire network; this, however, may not be feasible for large-size networks, as implementation of a highly reliable and fast communication system may be impractical or too costly. It is, therefore, necessary to design an feedback control law with a simple structure that requires access only to local information, while providing a satisfactory performance level. 

\subsection{Decentralized Feedback Control}\label{SEC_decentralized}

In this subsection, the objective is to design a static state-feedback control law with a one-hop information structure for problem (\ref{genPerfIndex}), (\ref{Gen_net}), (\ref{general_linear_cost}). Such a control law, for a general network, is of the form
\begin{equation}
u_i^k=\phi_i(x_i^k, (x_j^k)_{j\in\mathcal{D}_i}),
\end{equation}
where $\mathcal{D}_i$ denotes the set of all cells, excluding cell $i$, leaving/entering the downstream junction (head) of cell $i$. From the definition of $\mathcal{D}_i$, it follows that $\mathcal{E}_i^+\subseteq \mathcal{D}_i$; also, for any $i\in\mathcal{E}_{\text{off}}$,  $\mathcal{D}_i=\{\}$. For example, in Figure~\ref{fig_gen_net}, $\mathcal{D}_1=\{2,3\}$ and $\mathcal{D}_4=\{5,6\}$.

Design of a decentralized feedback controller can be viewed as solving an uncertain optimization problem, wherein non-local variables/parameters are unknown. The main challenges are how to ensure the feasibility of the solution and how to express or implement it in a feedback form.  

Uncertain linear program has been the subject of a lot of research and several approaches have been proposed to deal with robust optimization problems \cite{Bertsimas2006} including: solving the problem for nominal values of the unknown parameters and then performing sensitivity analysis; formulating the problem as a stochastic optimization by incorporating the knowledge on the probability distribution of the uncertain parameters; and assigning a finite set of possible values to the uncertain parameters and determining a solution which is relatively good for all the scenarios \cite{Gabrel2010}. Also, some research has focused on evaluating the impact of uncertainty on the cost by computing the worst and best optimum solutions \cite{Chinneck2000}. In some other works, in order to ensure the feasibility of solution, a worst-case approach is considered which, in general, leads to extremely conservative solutions \cite{Bertsimas2006}.

In this paper, we follow the decentralized procedure proposed in Section~\ref{SEC_controlDesign} which lead to a simple decentralized control law with the desired information structure and provides a feasible solution that under certain conditions could provide the optimal centralized performance. 

In order to design the $i$-th control law with a one-hope information structure, we design a centralized optimal static state-feedback controller for the sub-network consisting of cells $i$ and any $j\in\mathcal{D}_i$, with zero inflow rate to cell $i$. Then, the $i$-th local optimization is
\begin{align}
&\min_{u_i, (u_j)_{j\in\mathcal{D}_i}} \Big{\{}{\hat\varphi}_i(x_i^N, (x_j^N)_{j\in\mathcal{D}_i})\;+\label{localPerfIndexTRAFF}\\
&\;\;\;\;\;\;\;\;\;\;\;\;\;\;\;\;\;\;\;\;\;\;\;\;\; {\sum}_{k=0}^{N-1} \hat{\psi}_i^k(x_i^k, (x_j^k)_{j\in\mathcal{D}_i}, u_i^k, (u_j^k)_{j\in\mathcal{D}_i})\Big{\}}\nonumber\\
&\;\;\;\;\;\;\text{s.t.}\;(x_i, (x_j)_{j\in\mathcal{D}_i}, u_i, (u_j)_{j\in\mathcal{D}_i})\in{\hat{\Theta}}_i,\nonumber
\end{align}
where 
\begin{equation}
\begin{gathered}
{\hat\varphi}_i=\alpha_i^Nx_i^N + {\textstyle\sum}_{j\in\mathcal{D}_i} \alpha_j^Nx_j^N,\\
\hat{\psi}_i^k=\alpha_i^k x_i^k + \beta_i^ku_i^k + {\textstyle\sum}_{j\in\mathcal{D}_i}(\alpha_j^k x_j^k + \beta_j^ku_j^k).
\end{gathered}
\end{equation}
and the constraint set ${\hat{\Theta}}_i$ is defined by (\ref{Gen_net}) with $y_i^k=0$, $\forall k$, wherein any constraint involving non-local variables are relaxed.  

\begin{theorem}\label{thm_GEN_DENTTRAFF}
	Consider the local optimization (\ref{localPerfIndexTRAFF}), for a sub-network of cells $i$, $j\in\mathcal{D}_i$. The solution can be expressed as
	\begin{equation}\label{suboptimal_i}
	\begin{gathered}
	(\hat{u}_i^k)^* = \textit{pwa}^k_i(x_i^k, (x_{j}^k)_{j\in\mathcal{D}_i}).
	\end{gathered}
	\end{equation}
	which is a piecewise affine function on polyhedra of local state variables whose parameters can be computed off-line. Moreover, it satisfies all constraints in (\ref{Gen_net}) for any $k, i$.
\end{theorem}

\emph{Proof}: The proof is given in the Appendix.

We refer to (\ref{suboptimal_i}) as a ``sub-optimal decentralized control law with a one-hope information structure''. It should be highlighted that the separability property of the objective function and constraints has enabled us to simply construct a local version of the centralized optimization problem as (\ref{localPerfIndexTRAFF}).

The natural question that arises is how to evaluate the performance and sub-optimality level of the above decentralized control scheme. As mentioned earlier, in general, performance analysis of decentralized controllers is a very difficult task. Due to the lack of analytical tools, performance evaluation can be done through extensive numerical simulations. It should be noted that although the above decentralization procedure involves relaxations that may affect the conservativeness of the solution, under certain conditions, performance degradation due to decentralization is zero. It can be easily verified that if the conditions in Theorem~\ref{thm_GEN} are satisfied, solving the local optimization (\ref{localPerfIndexTRAFF}) gives the true optimal controller.   

\begin{example}\label{example01_3cell}
	To illustrate the decentralization process, let us consider a 3-cell network
	 \begin{figure}[h!]
	 	\centering
	 	\includegraphics[scale=0.3]{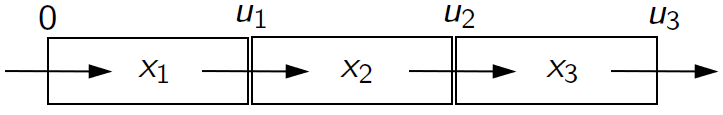}
	 	\caption{A 3-cell network with zero inflow rate.}
	 	\label{3cellExample}
	 \end{figure}
 with the following cost function and constraints
 \begin{equation}\label{example01}
 \begin{split}
&\min J={\textstyle\sum}_{k=0}^N(x_1^k + 4x_2^k+2x_3^k)\\
&\;\;\text{s.t.}\;\; x_{i}^{k+1}=x_i^k+u_{i-1}^k-u_i^k,\;\;\;u_0^k=0,\\
&\;\;\;\;\;\;\;\;0\leq u_i^k\leq 0.9x_i^k,\;\;\;u_i^k\leq1-0.3x_{i+1}^k.
 \end{split}
 \end{equation}
In a decentralized control with a one-hop information structure, given local state  at time $t$, the control action at time $t=0,1, \ldots, N-1$, is obtained as follows.

\vspace*{5pt}
\begin{center}
	\begin{tabular}{ p{4.3cm} | p{3.7cm} }
		\hline
		Local optimization at time $t$ & Action to be implemented\\\hline
		 \vspace*{0mm}$\min  J_1^t={\textstyle\sum}_{k=t}^N(x_1^k + 4x_2^k)$\newline
		 $x_{1}^{k+1}=x_1^k+0-u_1^k$,\newline
		 $x_{2}^{k+1}=x_2^k+u_1^k-u_2^k$,\newline
		 $0\leq u_1^k\leq 0.9x_1^k$, \newline
		 $u_1^k\leq1\!-\!0.3x_{2}^k$, \newline
		 $0\leq u_2^k\leq 0.9x_2^k$,
		& \vspace*{0mm}$u_1^t=\phi_1(x_1^t, x_2^t)$\\\hline
		\vspace*{0mm}$\min  J_2^t={\textstyle\sum}_{k=t}^N(4x_2^k + 2x_3^k)$\newline
		$x_{2}^{k+1}=x_2^k+0-u_2^k$,\newline
		$x_{3}^{k+1}=x_3^k+u_2^k-u_3^k$,\newline
		$0\leq u_2^k\leq 0.9x_2^k$, \newline
		$u_2^k\leq1-0.3x_{3}^k$, \newline
		$0\leq u_3^k\leq 0.9x_3^k$,
		& \vspace*{0mm}$u_2^t=\phi_2(x_2^t, x_3^t)$\\\hline
		\vspace*{0mm}$\min  J_3^t={\textstyle\sum}_{k=t}^N 2x_3^k$\newline
		$x_{3}^{k+1}=x_3^k+0-u_3^k$,\newline
		$0\leq u_3^k\leq 0.9x_3^k$,
		& \vspace*{0mm}$u_3^t=\phi_3(x_3^t)$\\\hline
	\end{tabular}
\end{center}

\vspace*{5pt}
Note that the cost weighting parameters in (\ref{example01}) do not satisfy conditions given in Theorem~\ref{thm_GEN}, so the uncontrolled solution may not be optimal. 
\end{example}

\section{Gas Network}\label{SEC_gas}

Consider a natural gas distribution network, where gas is transported through horizontal pipes. Flow regulation in gas networks is typically done by means of \emph{compressors} and/or \emph{control valves}. Transmission of natural gas over long distances and meeting consumers demand required the use of compressors to increase the pressure and to overcome pressure loss caused by friction in pipes. Also, in the distribution part of the network or where pipelines have low pressure limits, control valves are used to reduce pressure and restrict gas flow rate \cite{Koch2015}.

The problem of regulating flow in a gas network (e.g. to minimize the total supply cost) is often formulated as an optimization problem subject to \emph{nonlinear flow-pressure relations} \cite{Wong1968, Wong1968b, Wolf2000}. A controlled natural-gas pipeline system is partitioned into multiple cells as illustrated in Section~\ref{SEC_Intro}, and the gas flow dynamic for the $i$-th cell is described by (\ref{CTM001}), where $\rho_i^k$ is the \emph{gas mass density} [lb/ft] of cell $i$, $y_i^k$ [lb/sec] is the inflow rate to cell $i$, $u_i^k$ [lb/sec] is the outflow rate from cell $i$, and $\ell_i$ is the length [ft] of cell $i$. From the \emph{ideal gas law}, the pressure in cell $i$ is given by $x_i^k=\nu^2_0\rho_i^k$, where $\nu_0$ is a positive constant \cite{Zlotnik2015}. Then, one may express the gas flow dynamic in terms of internal pressures $x_i^k$ as $x_i^{k+1}=x_i^k+\tau_i(y_i^k-u_i^k)$, where $\tau_i$ is a positive constant dependent on the physical characteristics of the pipeline. 

If there is no regulator at the interface between cells $i$ and $i+1$ (passive interface), the flow rate from cell $i$ to the downstream cell $i+1$ satisfies the nonlinear equation $(u_i^k)^2 = \kappa_i^2 \left((x_i^k)^2-(x_{i+1}^k)^2\right)$, where $\kappa_i$ is a constant characterizing the pressure drop due to flow in the pipe which is dependent upon the pipe physical characteristics (diameter, length, rugosity) \cite{Wolf2000}. In the presence of a regulator (active interface), however, the flow and pressures satisfy $(u_i^k)^2 \geq \kappa_i^2 \left((x_i^k)^2-(x_{i+1}^k)^2\right)$ for a \emph{compressor}, and $(u_i^k)^2 \leq \kappa_i^2 \left((x_i^k)^2-(x_{i+1}^k)^2\right)$ for a \emph{control valve}.

For the sake of simplicity, we focus on gas networks with a line structure as shown in Figure~\ref{fig001}, where cells are increasingly numbered from upstream to downstream. In addition, we assume that the gas flow is regulated by \emph{control valves} such that all pressure/flow constraints are met. We consider a cost function of the form
\begin{equation}\label{GasCost001}
\begin{gathered}
J= {\sum}_{k=0}^{N-1}\beta^ku_0^k + {\sum}_{k=0}^N {\sum}_{i=1}^n\alpha_i^k(x_i^k-x_{i+1}^k),
\end{gathered}
\end{equation}
where $u_0^k$ is the exogenous inflow to the network coming from the high-pressure part of the network, $\beta^k\geq 0$ is the purchase price of the gas, $x_i^k-x_{i+1}^k$ is pressure drop in the $i$-th control valve, and  $\alpha_i^k\geq 0$ is a cost-weighting parameter. The objective is to minimize the cost in (\ref{GasCost001}), subject to
\begin{subequations}\label{GenGasNet}
	\begin{align}
	&x^{k+1}_i=x^k_i+\tau_i(u^k_{i-1}-u^k_i),\;\;\;\forall i\in\mathbb{N}_n \label{GenGasNet_a}\\
	&(u_i^k)^2 \leq \kappa_i^2 \left((x_i^k)^2-(x_{i+1}^k)^2\right), \label{GenGasNet_b}\\
	&0\leq x_i^k - x^k_{i+1}\leq \Delta_i,  \label{GenGasNet_c}\\
	&0\leq \underline{u}_i\leq u_i^k\leq \bar{u}_i,\\
	&0< \underline{x}_i\leq x_i^k\leq \bar{x}_i, \label{GenGasNet_e}
	\end{align}
\end{subequations}
where $\underline{u}_i, \bar{u}_i$ and $\underline{x}_i, \bar{x}_i$ are known limits on the flow rate and pressure of cell $i$, $\Delta_i$ is a known upper limit for pressure drop in the $i$-th control valve, $u_n^k$ is the outflow rate (consumer demand flow) which is known over the control horizon, $x_{n+1}^k$ denotes the output pressure of the network, and control variables are $u_i^k$, for $i=0,1,\ldots, (n-1)$ and $k=0,1,\ldots, (N-1)$.

Unlike the highway traffic problem (\ref{genPerfIndex}), (\ref{Gen_net}), (\ref{general_linear_cost}) which is a linear program (LP), the optimization problem (\ref{GenGasNet}) has nonlinear inequality constraints (\ref{GenGasNet_b}), making it non-convex and difficult to design a globally optimal feedback solution. There are  numerous convexification/approximation techniques to deal with non-convex nonlinear programs (NLPs). Linear approximation, for instance, is a standard and widely-used technique in nonlinear programming. In order to find an approximate solution to (\ref{GasCost001}), (\ref{GenGasNet}) in a feedback form, we convert it into an LP. Two approximations techniques are considered:
\begin{itemize}
	\item \textbf{LP1}: A convexification of (\ref{GenGasNet_b}) that makes it linear and guarantees feasibility of the approximate optimizer with respect to the original problem
	\item \textbf{LP2}: Through piecewise-affine approximation of the nonlinear terms in (\ref{GenGasNet_b}), an approximate solution is obtained that provides an approximation to the global optimum cost value which can be made arbitrarily close to the true global optimum cost value by improving the approximation quality (increasing the number of breakpoints). The approximate optimizer, however, is not guaranteed to be feasible.   
\end{itemize}
We use the solution to LP1 for implementation (which satisfies all the constraints) and use the optimum cost value of LP2 to evaluate the level of sub-optimality of the solution to LP1. Procedures for constructing LP1 and LP2 are described below.     

We consider the following convexification to make LP1. From (\ref{GenGasNet_c}), (\ref{GenGasNet_e}), pressures satisfy $x_i^k\geq x_{i+1}^k>0$, then the relation
$(x_i^k-x_{i+1}^k)^2\leq (x_i^k)^2-(x_{i+1}^k)^2$ holds. Then, in (\ref{GenGasNet_b}), the upper limit of $(u_i^k)^2$  can be replaced by a more conservative bound, i.e. $(u_i^k)^2 \leq \kappa_i^2(x_i^k-x_{i+1}^k)^2$, or equivalently $u_i^k\leq \kappa_i(x_i^k-x_{i+1}^k)$, reducing (\ref{GasCost001}), (\ref{GenGasNet}) into an LP whose solution can be expressed in a feedback form (see Theorem~\ref{thm0}). The feasibility of solution to LP1 is guaranteed as the constraint set of LP1 is a subset of that of the original problem.

In order to assess the sub-optimality of the solution to LP1, we construct LP2 as follows.  Since the nonlinear terms in (\ref{GenGasNet_b}) are mathematically independent, one may replace each nonlinear term with a piecewise-affine approximation and utilize \emph{separable programming} technique \cite{Stefanov2001}. The linearization procedure and the underlying assumptions are summarized below.

\begin{algorithm}\cite{Jensen2003}[\S 10.4]\label{alg01} (\emph{Piecewise Linear Approximation of a Separable Nonlinear Program}) Consider the problem of minimizing a real-valued function $f(z)$ over a closed and bounded region $\Omega=\{z\in\mathbb{R}^n\,|\,g_j(z) \leq b_j,\, j\in\mathbb{N}_q\}$, and assume that the following properties hold: 
\begin{itemize}
	\item [(i)] Every decision variable is bounded from below and above by known constants, i.e. $\underline{z}_i\leq z_i\leq \bar{z}_i$, $\forall i\in\mathbb{N}_n$.
	\item [(ii)] Decision variables appear \emph{separately} in the cost and constraints, i.e. $f$ and $g_j$, $\forall j\in\mathbb{N}_q$, can be expressed as a sum of functions of a single variable: $f(z)={\textstyle\sum}_{i=1}^{n}f_{i}(z_i)$ and  ${\textstyle\sum}_{i=1}^{n}g_{ji}(z_i)\leq b_j$.
\end{itemize}
Then, the NLP $\min_{z\in\Omega}f(z)$ in the $z$-space is transformed into an \emph{approximate linear program} (ALP) in $\xi$-space as follows: For each decision variable $z_i\in\mathbb{R}$, consider $m_i$ breakpoints (not necessarily equally spaced) $\underline{z}_i=z_{i1}\leq z_{i2}\leq \ldots \leq z_{im_i}= \bar{z}_i$, then express each decision variable in terms of breakpoints as 
\begin{equation}
z_i={\textstyle\sum}_{j=1}^{m_i}\xi_{ij}z_{ij} = \xi_i^\top \hat{z}_i,\;\;\;\;{\textstyle\sum}_{j=1}^{m_i}\xi_{ij}=\xi_i^\top\mathbf{1}=1,
\end{equation}
where $\xi_{ij}\geq 0$, $\xi_i=[\xi_{i1}, \ldots, \xi_{im_i}]^\top$, $\mathbf{1}$ is column vector of ones, and $\hat{z}_i=[z_{i1}, \ldots,$ $z_{im_i}]^\top$ is the vectors of breakpoints for variable $z_i$. Therefore, in the $\xi$-space, the cost function and constraints can be written as
\begin{equation*}
f(\xi)={\textstyle\sum}_{i=1}^{n}{\textstyle\sum}_{k=1}^{m_i}\xi_{ik}f_{i}(z_{ik})=
{\textstyle\sum}_{i=1}^{n} \xi_i^\top \hat{f}_i
\end{equation*}
\begin{equation*}
{\textstyle\sum}_{i=1}^{n}{\textstyle\sum}_{k=1}^{m_i}\xi_{ik}g_{ji}(z_{ik}) = 
{\textstyle\sum}_{i=1}^{n} \xi_i^\top \hat{g}_{ji} \leq b_j,\;\;\forall j\in\mathbb{N}_q,
\end{equation*}
where $\hat{f}_i=[f_i(z_{i1}), \ldots, f_i(z_{im_i})]^\top$ and $\hat{g}_{ji}=[g_{ji}(z_{i1}), \ldots,$ $g_{ji}(z_{im_i})]^\top$ are respectively vectors of $f_i(z_i)$ and $g_{ji}(z_i)$ evaluated at the breakpoints of $z_i$. The approximation quality can be improved arbitrarily by increasing the number of breakpoints or by a suitable selection of  breakpoints distribution for each nonlinear term. $\hfill\square$
\end{algorithm}

The important questions in connection with Algorithm~\ref{alg01} are: (i) whether the solution to an ALP is feasible with respect to the original problem; and (ii) if a solution to the ALP is an approximate global (not local) optimum to the original problem. The following lemma states that the answer to the both questions is yes, if the original problem is convex.

\begin{lemma}\cite{Stefanov2001, Jensen2003}\label{lem_0002a}
	Consider a separable NLP as described in Algorithm~\ref{alg01}, ans assume that $f(z)$ and $g_j(z)$, $\forall j$, are separable \emph{convex} functions, for some convex functions $f_i$ and $g_{ji}$. An  optimal solution to an ALP (obtained using any standard LP solver) gives a feasible approximation to global optimizer of the original problem. Moreover, the accuracy of the solution is arbitrarily improved by increasing the number of breakpoints $m_i$ for each decision variable $z_i$.  
\end{lemma}	 

It should be noted that in Lemma~\ref{lem_0002a}, there is no need to enforce any modification such as \emph{adjacency restriction} (see \cite{Stefanov2001},  \cite{Jensen2003}[\S 10.4]) to standard LP solvers when solving the ALP. In general, however, applying separable programming (with or without adjacency restriction) to a non-convex problem may give an approximation to a local optimizer or an infeasible solution with respect to the original problem.
  

The following lemma gives some properties of \emph{concave minimization}. These properties are useful to determine the optimal cost value for the gas network problem (\ref{GasCost001}), (\ref{GenGasNet}), wherein the cost function is \emph{linear} (hence both convex and concave).

\begin{lemma}\label{lem_0001}
	Consider the problem of minimizing a real-valued continuous \emph{concave} function $f$ over a closed and bounded (possibly non-convex) region $\Omega\subset\mathbb{R}^n$. Then, (i) $f$ attains its global minimum over any closed and bounded set; (ii) $\min \{f(z)\;|\;z\in\Omega\} = \min \{f(z)\;|\;z\in\text{conv}(\Omega)\}$, where $\text{conv}(\Omega)$ denotes the \emph{convex hull} of $\Omega$; (iii) $f$ attains its global minimum at an extreme point of $\text{conv}(\Omega)$.
\end{lemma}	

\emph{Proof}: (i) Since $f$ is a continuous function, then the existence of a global minimizer over any compact (closed and bounded) set follows from the well-known Weierstrass Theorem. For the proof of (ii) and (iii), see \cite[Proposition~2.3]{Hager2016} and \cite[Theorem~32.2]{Rockafellar70}, and \cite[Theorem~1.19]{Horst2000}.  $\hfill\blacksquare$

From Algorithm~\ref{alg01} and Lemma~\ref{lem_0001}, it follows that solving an ALP associated with problem (\ref{GasCost001}), (\ref{GenGasNet}) using any standard LP solver (e.g. standard simplex method with no modification) provides an approximation to the global optimum cost value; moreover, the approximation error is arbitrarily reduced by increasing the number of breakpoints. The optimizer, however, may not be feasible as it belongs to $\text{conv}(\Omega)$ not $\Omega$.

\subsection{Centralized Feedback Control}\label{SEC_centralizedGAS}

As mentioned earlier, in order to design a centralized \emph{feedback control law} for the optimal control problem (\ref{GasCost001}), (\ref{GenGasNet}), we replace  (\ref{GenGasNet_b}) with $u_i^k\leq \kappa_i(x_i^k-x_{i+1}^k)$ and solve the LP by following the same procedure explained in Section~\ref{SEC_traffic}. Hence, the sub-optimal solution is expressed as
$$(u^k)^* =\textit{pwa}^k(x^k).$$
which is guaranteed to be feasible but may bot be truly optimal. To evaluate the level of sub-optimality of the solution we solve the corresponding LP2 problem.  The linearity of the cost function and separability of the nonlinear constraints (\ref{GenGasNet_b}) ensures that solving an ALP gives an approximation to the global optimum cost value. Comparing the optimal cost values of LP1 and LP2 gives the error introduced due to convexification in LP1.  

\begin{remark}
	For any convex problem, applying LP2 gives a feasible approximate to global optimum; hence, for flow networks with any convex objective function and constraint set one can design an optimal controller in a feedback form.
\end{remark}

\subsection{Decentralized Feedback Control}\label{SEC_decentralizedGAS}

By following the same procedure as that is Section~\ref{SEC_decentralized}, we can design a decentralized control law with a one-hop information structure for problem (\ref{GasCost001}), (\ref{GenGasNet}), we follow  and solve the LP1 associated with each sub-network and convexification errors can be found by solving the corresponding LP2. 

From (\ref{GenGasNet}), the feasibility of $u_i^k$ depends on the values of $x_i^k$ and $x_{i+1}^k$ and is independent of the state of non-local variables. Hence, in a decentralization feedback control law with a one-hop structure, the feasibility of $u_i^k$ is ensured in the $i$-th local optimization problem. 


\section{Numerical Simulation}\label{SEC_simulation}

As mentioned earlier, it is generally difficult to analytically evaluate performance of a decentralized control scheme in compared with that of an optimal centralized controller, hence comprehensive numerical simulations must be performed to demonstrate the effectiveness of a decentralization technique and and numerically assess the level of sub-optimality.

\begin{figure}[h!]
	\centering
	\includegraphics[scale=0.3]{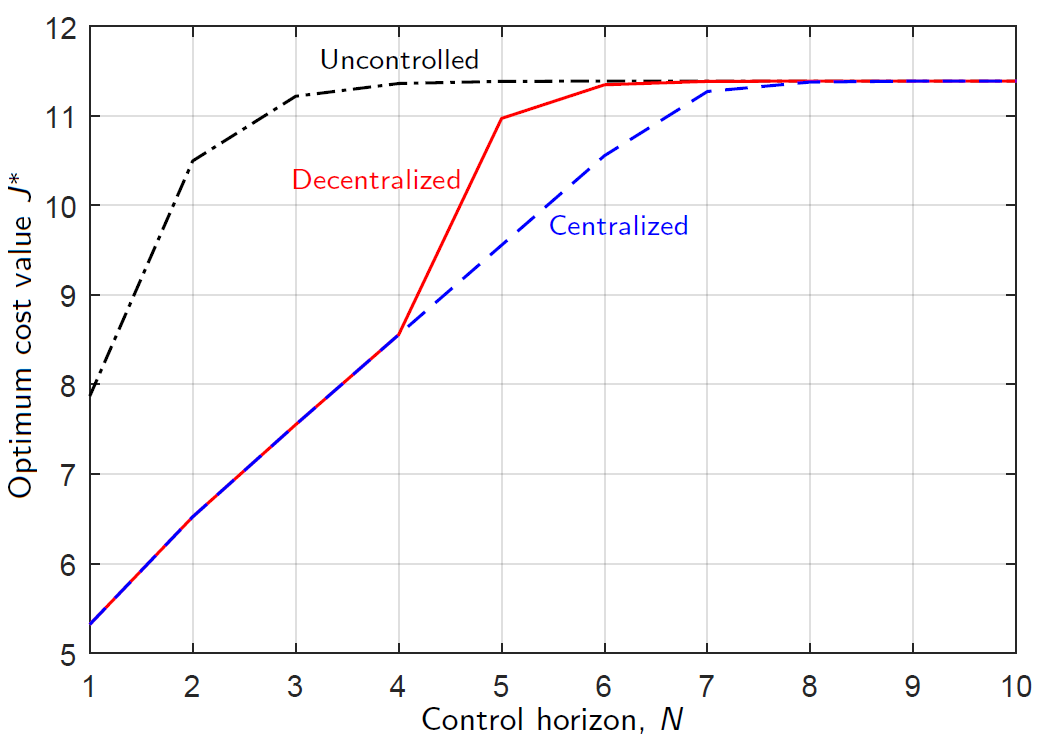}
	\caption{Performance of three control schemes as a function of control horizon for the 3-cell network in Example~\ref{example01_3cell}.}
	\label{sim1Plots}
\end{figure}

\emph{Simulation~1}: Consider the 3-cell network in Example~\ref{example01_3cell}, with initial state $x^0=[1, 0.5, 0.1]$. Figure~\ref{sim1Plots} shows the optimal cost value for different values of control horizon $N$, for three schemes: (i) Uncontrolled, i.e. at each time setting every control variable to its maximum possible value, (ii) Decentralized control with one-hop information structure, and (iii) Centralized control. In this example, for short control horizons $N=1,2,3,4$, there is no performance loss due to decentralization. For $N=5$, there a $\%15$ relative error between the performance of centralized and decentralized schemes. For larger values of $N$, the performance of the three schemes converge to each other (for $N\geq 8$, the network is almost evacuated).   
It should be noted that the uncontrolled scheme (trivial control) has a decentralized form with a one-hop information structure, but it is obtained by setting each decision variable to its maximum allowed value involving no local optimization and computational cost. 

\begin{figure}[h!]
	\centering
	\includegraphics[scale=0.30]{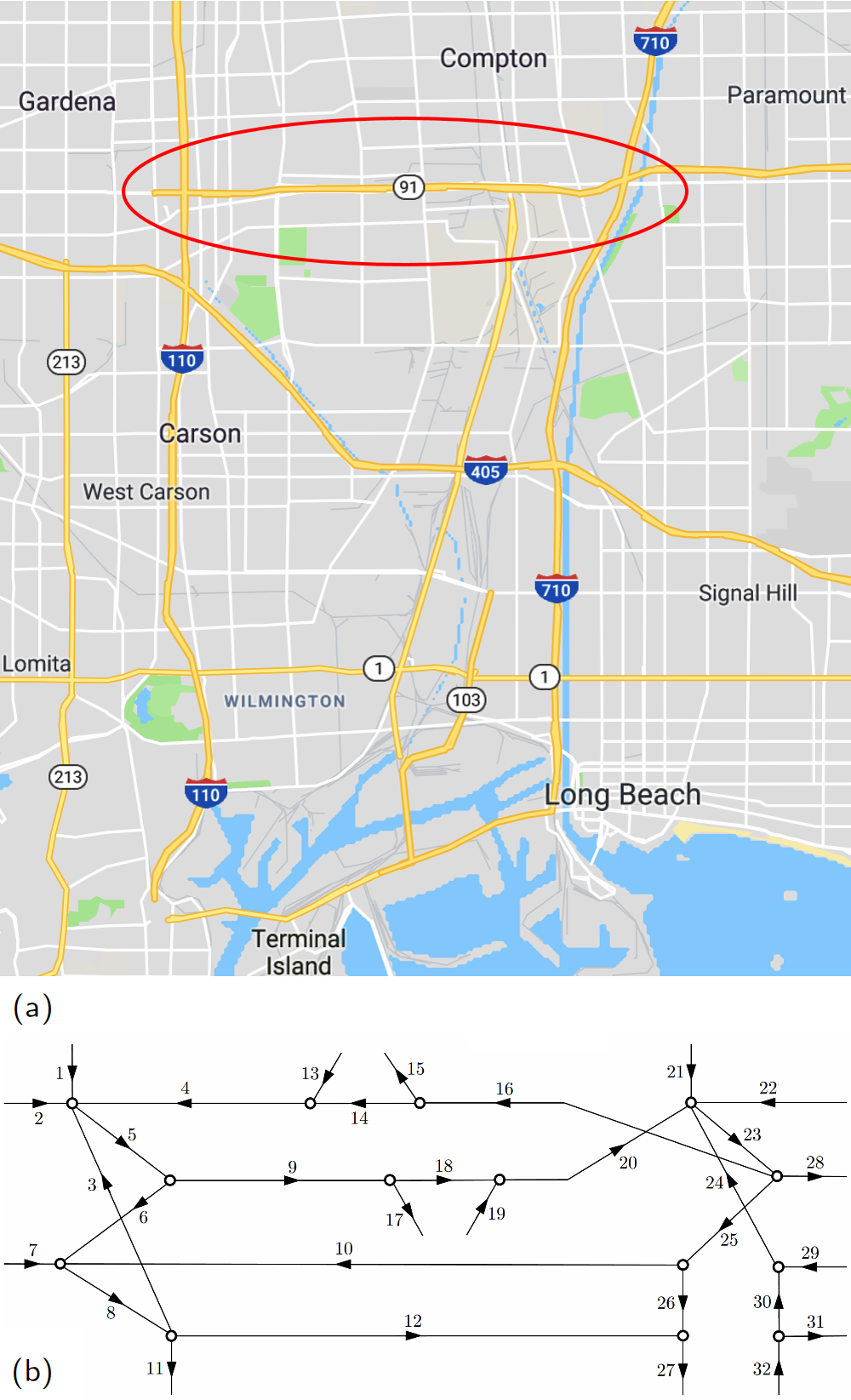}
	\caption{(a) The map of an area in the southern Los Angeles. The red ellipse shows the region used in our numerical simulation. (b) The directed graph of the transportation network of the region of interest with 32 cells, where $\mathcal{E}_{\text{on}}=\{1, 2, 7, 13, 19, 21, 22, 29, 32\}$ and $\mathcal{E}_{\text{off}}=\{11, 15, 17, 27, 28, 31\}$. }
	\label{example_32_MAP}
\end{figure}

\begin{figure}[h!]
	\centering
	\includegraphics[scale=0.36]{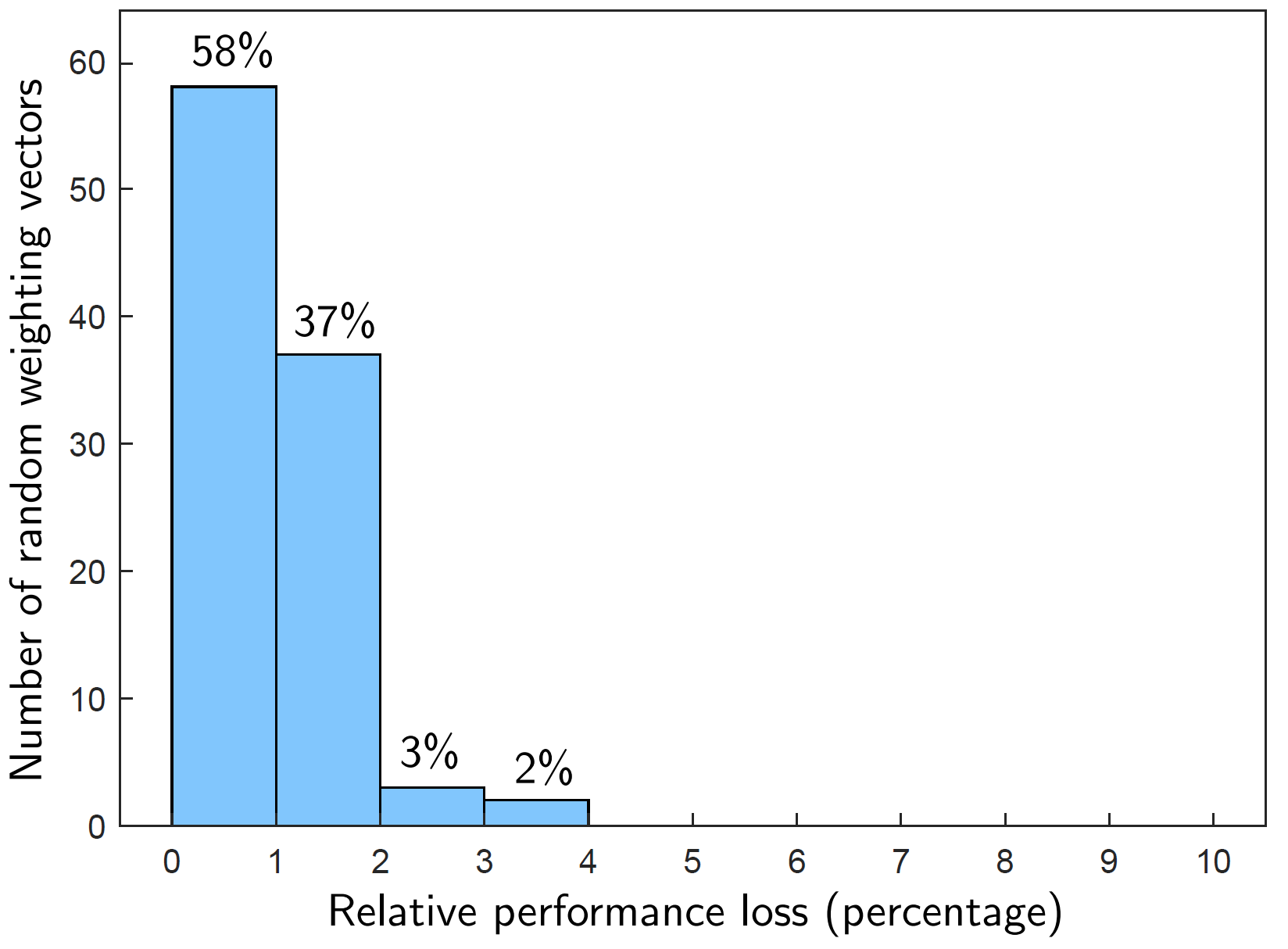}
	\caption{The histogram of the relative decentralization performance loss for $100$ random cost-weighting vectors.}
	\label{example_32cell_perf}
\end{figure}

\emph{Simulation~2}: To evaluate the performance of the decentralized scheme, let us consider a larger network with more realistic architecture and parameters. We consider the freeway system of an area in the southern Los Angeles as shown in Figure~\ref{example_32_MAP}(a) modeled by the CTM. The directed graph of the network of the region of interest consisting of 32 cells is shown in Figure~\ref{example_32_MAP}(b).
Consider minimization of  $J={\textstyle\sum}_{k=0}^{20}{\textstyle\sum}_{i=1}^n \alpha_i x_i^k$, subject to (\ref{Gen_net}), with the following parameters: The sampling time is $T_s= 1/360$~hr (or $10$~sec). For on-ramp cells, the jam traffic density $\gamma_i$ is assumed to be infinity and for other cells $\gamma_i=200$~veh/mi. For all cells, the backward congestion wave traveling speed is $w_i=13$~mi/hr. For cells $3, 4, 9, 10, 12, 16, 20$, the cell's length is $\ell_i=2$~mi, the free-flow speed is $v_i=65$~mi/hr, and the maximum flow capacity is $C_i=800$~veh/hr, and for other cells, $\ell_i=0.5$~mi, $v_i=25$~mi/hr, and $C_i=400$~veh/hr. At any diverge junction, $\hbar_i$ with incoming cell $i$, the turning ratios are time-invariant and are split uniformly between the outgoing cells, i.e., $R_{ij}=1/n_{\hbar_i}$, where $n_{\hbar_i}$ is the number of outgoing cells from junction $\hbar_i$.  Let $\alpha=[\alpha_1, \alpha_2, \ldots, \alpha_{32}]$ be the cos weighting vector, where $\alpha_i$ is the weight associated with the state of cell $i$ in cost function. We assign random integers between $1$ and $6$ to $\alpha_i$'s and compute the optimal cost value of the centralized controller $J^*_{\text{cen}}$ and that of the decentralized one (with one-hop information structure) $J^*_{\text{dec}}$, and then evaluate the relative decentralization performance loss $\varepsilon=100(J^*_{\text{dec}}-J^*_{\text{cen}})/J^*_{\text{cen}}$. For example, with $\alpha=[5, 1,$ $ 2, 2, 4, 1, 3, 1, 1, 5, 2, 4, 6, 3, 5, 5, 3, 4, 1, 6, 2, 2, 5, 3, 5, 3, 2, 1,$ $5, 3, 3, 4]$, we have $\varepsilon=0.994\%$. We consider $100$ random weighting vectors, $\alpha$, and for each case evaluate the relative performance loss $\varepsilon$. Figure~\ref{example_32cell_perf} shows the histogram of the relative errors, wherein for $95\%$ of weighting vectors, the relative decentralization performance loss is less than $2\%$.


\section{Conclusion}\label{SEC_conclusion}
This paper provides some structural insights into the finite-horizon optimal feedback control for flow networks. The enabling tool for the design of an optimal feedback control law is the multi-parametric linear program. It is well known that for large-size complex networks, the prohibitive computation and computation loads makes the design and implementation of a centralized controller too costly or impractical; moreover, the effect of noise, delay, or any type of error or failure in data transmission may substantially degrade the control quality. It is, therefore, necessary to develop decentralized feedback controllers with simple structure. A simple procedure is proposed to design a decentralized feedback control with a ``one-hop'' information structure.  Moreover, it is shown that the optimal feedback controller with respect to certain linear performance indexes possesses a one-hop information structure, making the optimal controller suitable for practical implementations in large-scale networks. This suggests that if certain conditions are satisfies, the \emph{trivial control} (with the least computational/communication cost) can provide the same (or very close) performance to that of the \emph{centralized control} (with the most computational/communication cost).  

For a given flow network of size $n$ and control horizon $N$, it is invaluable to analytically determine when it is worth to implement uncontrolled scheme, or a decentralized control law with a $p$-hop information structure to achieve a satisfactory level of permanence. Our ultimate objective is to develop a principled approach for distributed optimal control of physical infrastructure networks under given information constraints.

\section*{Appendix}

\emph{Proof of Theorem~\ref{thm1}}:
The proof follows by following a similar procedure as that in \cite[\S 2]{Borrelli_Thesis_2002}. The closed-form solution to (\ref{Gen_net_a}) starting from initial state $x^0$ is given by
\begin{equation}\label{closed-form_state_eq}
\begin{split}
&x^{k}_i=x^0_i+T_s{\textstyle\sum}_{j=0}^{k-1}(y^j_i-u^j_i),\;i\in\mathbb{N}_n,
\end{split}
\end{equation}
where from (\ref{Gen_net_b}), the total inflow rate to cell $i$ at time $j$ is
\begin{equation}\label{tot_inflow}
\begin{split}
y_i^j &= \begin{cases}
\lambda_i^j, & \text{if } i\in\mathcal{E}_{\text{on}}\\
{\textstyle\sum}_{q=1}^nR_{qi}u_q^j, & \text{if } i\not\in\mathcal{E}_{\text{on}}
\end{cases},\\
&= \begin{cases}
\lambda_i^j, & \text{if } i\in\mathcal{E}_{\text{on}},\\
R_{qi}u_q^j, & \text{if } i\not\in\mathcal{E}_{\text{on}} \text{ and }\tau_i \text{ is either diverge}\\
&\hfill\text{or ordinary, with } \mathcal{E}_{i}^-\!=\!\{q\},\\
{\textstyle\sum}_{q\in\mathcal{E}_{i}^-}u_q^j, & \text{if } i\not\in\mathcal{E}_{\text{on}} \text{ and }\tau_i \text{ is merge,}
\end{cases}
\end{split}
\end{equation}
where $\tau_i$ denotes the tail (upstream junction) of cell $i$. Then, from (\ref{genPerfIndex}), (\ref{general_linear_cost}), (\ref{closed-form_state_eq}), and (\ref{tot_inflow}), the cost function can be expressed as a linear combination of $u_i^k$'s as follows:
\begin{equation*}
\begin{split}
J &= \sum_{i=1}^n\sum_{k=0}^N\alpha_i^kx_i^0 + T_s\sum_{i=1}^n\sum_{k=1}^N\sum_{j=0}^{k-1}\alpha_i^ky_i^j\\
&\;\;-
T_s\sum_{i=1}^n\sum_{k=1}^N\sum_{j=0}^{k-1}\alpha_i^ku_{i}^j +
\sum_{i=1}^n\sum_{k=0}^{N-1}\beta_i^ku_i^k\\
&= \sum_{i=1}^n\sum_{k=0}^N\alpha_i^kx_i^0 + T_s\sum_{k=1}^N\sum_{j=0}^{k-1}\sum_{i\in\mathcal{E}_{\text{on}}}\alpha_i^k\lambda_i^j\\
&\;\;-T_s\sum_{k=1}^N\!\sum_{j=0}^{k-1}\!\Big{(}\!\sum_{i=1}^n\!\!\alpha_i^ku_{i}^j \!-\!\!\sum_{i\not\in\mathcal{E}_{\text{on}}}\!\sum_{q=1}^n\!\!\alpha_i^kR_{qi}u_q^j\Big{)}\!+\!\sum_{i=1}^n\!\sum_{k=0}^{N-1}\!\!\beta_i^ku_i^k\\
&= \sum_{i=1}^n\sum_{k=0}^N\alpha_i^kx_i^0 + T_s\sum_{k=1}^N\sum_{j=0}^{k-1}\sum_{i\in\mathcal{E}_{\text{on}}}\alpha_i^k\lambda_i^j\\
&\;\;-T_s\sum_{q=1}^n\sum_{k=1}^N\sum_{j=0}^{k-1}\Big{(}\alpha_q^k -\sum_{i\not\in\mathcal{E}_{\text{on}}}\alpha_i^kR_{qi}\Big{)}u_q^j+\sum_{i=1}^n\sum_{k=0}^{N-1}\beta_i^ku_i^k.
\end{split}
\end{equation*}
Then, the cost function over horizon $[0, N]$ can be written as
\begin{equation}\label{new_cost_gen001}
J^{[0,N]}= g(x^0, \lambda)-\!\sum_{k=0}^{N-1}\left(\mu_1^ku_1^k +\mu_2^ku_2^k+\ldots+\mu_n^ku_n^k\right),
\end{equation}
where 
\begin{equation}
\begin{split}
g(x^0,\lambda)=\sum_{i=1}^n\sum_{k=0}^N\alpha_i^kx_i^0 + T_s\sum_{k=1}^N\sum_{j=0}^{k-1}\sum_{i\in\mathcal{E}_{\text{on}}}\alpha_i^k\lambda_i^j,
\end{split}
\end{equation}
and the coefficients of the decision variables are given by
\begin{equation}
\begin{split}
\mu_q^k=-\beta_q^k+ {\sum}_{j=k+1}^NT_s\big{(}\alpha_q^j-{\sum}_{i\not\in\mathcal{E}_{\text{on}}}R_{qi}\alpha_i^j\big{)}.
\end{split}
\end{equation}

From (\ref{Gen_net_b}) and (\ref{tot_inflow}), if the head (downstream junction) of cell $i$ is either ordinary or diverge, the constraint for $u_i^k$ is 
\begin{equation}\label{case_1_junction}
\begin{gathered}
0\leq u_i^k\leq \bar{u}_i^k,\\
\bar{u}_i^k \!=\! \min\left\{\!\frac{v_i}{\ell_i}x_i^k, C_i,\Big{(}  \frac{w_{j}}{R_{ij}}(\gamma_{j}\!-\!\frac{1}{\ell_{j}}x_{j}^k), \frac{C_{j}}{R_{ij}}\Big{)}_{j\in\mathcal{E}_i^+}\!\right\},
\end{gathered}
\end{equation}
and if the head of cell $i$ is a merge junction with $\mathcal{E}_i^+=\{q\}$, then $u_i^k$ must satisfy
\begin{equation}\label{case_2_junction}
\begin{split}
&0\leq u_i^k \leq \min\left\{\frac{v_i}{\ell_i}x_i^k, C_i\right\},\;\;\forall i\in\mathcal{E}_q^-,\\
&{\sum}_{i\in\mathcal{E}_q^-}u_i^k \leq \min\left\{w_q\Big{(}\gamma_q-\frac{1}{\ell_q}x_q^k\Big{)}, C_q\right\}.
\end{split}
\end{equation}  

Due to the linearity of the objective function (\ref{new_cost_gen001}) and constraints (\ref{case_1_junction})-(\ref{case_2_junction}), the optimization problem can be expressed as a multi-parametric linear program of the form (\ref{mp_LP}), wherein the state vector at time $k=0$, i.e., $\theta=x^0$, is treated as a varying parameter vector in the optimization problem, and the decision variable vector contains the control actions for $k=0,\ldots, N-1$, i.e., $z=[(u^0)^\top, \ldots, (u^{N-1})^\top]^\top$. From Theorem~\ref{thm0}, we have
$$z^*={L}_{i}x^0 + {l}_{i},\;\;\text{if }\Pi_ix^0\leq \eta_i,$$
which can be expressed as 
\begin{equation}\label{open_0_N}
\begin{cases}
(u^0)^*={L}_{0i}x^0 + {l}_{0i},\\
(u^1)^*={L}_{1i}x^0 + {l}_{1i}\\
\;\;\;\;\;\;\;\;\;\,\vdots\\
(u^{N-1})^*={L}_{(N-1)i}x^0 + {l}_{(N-1)i}
\end{cases}\!\!\!\!\!\!,
\text{if } \Pi_ix^0\leq \eta_i,
\end{equation}
where $L_{ji}$ is the $j$-th row of matrix  $L_i$ and $l_{ji}$ is the $j$-th element of vector $l_i$. The above results imply that when optimizing the performance index starting at $k=0$ over the control horizon $[0, N]$, i.e., $J^{[0,N]}$ with parameter vector $\theta=x^0$, then the optimal solution (\ref{open_0_N}) provides a static state-feedback optimal control law only at the initial time $k=0$, i.e., 
$$(u^0)^*={L}_{0i}x^0 + {l}_{0i},\;\;\text{if }\Pi_ix^0\leq \eta_i.$$ 
Hence, to design a feedback control law, we retain only the first equation in (\ref{open_0_N}) and discard the rest of them. Therefore, in (\ref{centralized_controller}), the parameters of the optimal feedback controller at time $k=0$ are given by
\begin{equation}
F_i^0=L_{0i},\;\;f_i^0=l_{0i},\;\;H_i^0=\Pi_i,\;\;h_i^0=\eta_i.
\end{equation}
The optimal value of $u^0$ when is applied to the system gives an optimal value of $x^1$, then at the next time step by repeating the same procedure starting at the initial time $k=1$ over the control horizon $[1,N]$ with $x^1$ as a parameter vector, we can  express the optimal value of $u^1$ as a piecewise affine function of $x^1$. Therefore, in general, optimizing the performance index over the time interval $[j, N]$, i.e., $J^{[j,N]}$ with parameter vector $\theta=x^j$ and decision variable vector $z=[(u^j)^\top, \ldots, (u^{N-1})^\top]^\top$, provides a state-feedback optimal control law at time step $j$ in the form of a piecewise affine function on polyhedra of $x^j$, for any $j=0, 1, \ldots, N-1$. Hence, by solving $N$  multi-parametric linear programs, the optimal feedback controller can be expressed as (\ref{centralized_controller}). $\hfill\blacksquare$

\emph{Proof of Corollary~\ref{cor01}}: The proof of the first part immediately follows from Theorem~\ref{thm1}. The state of an on-ramp is given by $x_i^k=x_i^0+T_s{\textstyle\sum}_{j=0}^{k-1}(\lambda_i^j-u_i^j)$. Since only the state of an on-ramp cell directly depends on exogenous inflow rates, we need to just check the constraints that depend on $x_i^k$ for $i\in\mathcal{E}_{\text{on}}$. From (\ref{Gen_net}), the only constraint depending on the state of on-ramps is $u_i^k\leq (v_i/\ell_i)x_i^k$, $\forall i\in\mathcal{E}_{\text{on}}$, which can be written as $u_i^k+(v_i/\ell_i)T_s{\textstyle\sum}_{j=0}^{k-1}u_i^j\leq (v_i/\ell_i)x_i^0+(v_i/\ell_i)T_s{\textstyle\sum}_{j=0}^{k-1}\lambda_i^j$. Since the last term on the right side of the inequality is non-negative, the feasibility of zero-$\lambda$ optimal solution is guaranteed, when it is applied to a network with nonzero $\lambda_i^k$.  $\hfill\blacksquare$

\emph{Proof of Theorem~\ref{thm_GEN}}: From (\ref{new_cost_gen001}) and (\ref{case_1_junction}), for a network with no merge junction we can write
\begin{equation}\label{cost_genhh6t79}
u^* \!= \argmax_{\text{s.t. }(\ref{case_1_junction})}\left\{\sum_{k=0}^{N-1}\!\left(\mu_1^ku_1^k +\mu_2^ku_2^k+\ldots+\mu_n^ku_n^k\right)\!\right\}.
\end{equation}
From the definition of split ratios, ${\sum}_{i\not\in\mathcal{E}_{\text{on}}}R_{qi}\alpha_i^j$ is a convex combination of $\alpha_i^j$'s, for all $i\in\mathcal{E}_q^+$, and under the assumptions on the cost-weighting parameters, i.e., $\alpha_q^k\geq \alpha_{i}^k\geq 0$, $\forall k, q$ and $\forall i\in\mathcal{E}_q^+$, and $\beta_q^k\leq \beta_q^{k+1}\leq 0$, $\forall k, q$, and that the split ratios are time invariant, it follows that $\mu_q^k\geq \mu_q^{k+1}\geq 0$, $\forall q, k$. 

By using the \emph{dynamic programming} approach \cite[\S 6.2]{Frank_Lewis12}, we show that under the above assumptions, an optimal solution is obtained when each decision variable $u_i^k$ is set equal to its upper limit $\bar{u}_i^k$ defined in (\ref{case_1_junction}). For the maximization problem (\ref{cost_genhh6t79}), we define the objective functional over the time interval $[k, N]$ starting at time $k$ from initial state $x^k$ as
\begin{equation}\label{obj_over_k_N}
I^k(x^k)={\sum}_{j=k}^{N-1}(\mu_1^ju_1^j +\ldots+\mu_n^ju_n^j).
\end{equation}
Then, the corresponding \emph{functional equation of dynamic programming} is given by
\begin{equation}\label{Q_function_GEN}
\begin{gathered}
{I^{k}}^*\!(x^k)=\max_{u^k}\,\{Q^k\},\;\;\text{s.t. }(\ref{case_1_junction}),\;\;\text{given }x^k,\\
Q^k=\left(\mu_1^ku_1^k +\ldots+\mu_n^ku_n^k\right) + {I^{k+1}}^*\!(x^{k+1}),
\end{gathered}
\end{equation}
where ${I^{k}}^*\!(x^k)$ denotes the \emph{optimal value} of the objective function over the time interval $[k, N]$ from initial state $x^k$. Hence, the optimization over the horizon $[0,N]$ is converted into optimization over only one control vector $u^k$ at a time by working \emph{backward in time} for $k=N-1, N-2, \ldots, 0$. The optimization problem (\ref{Q_function_GEN}) is a \emph{bound constrained optimization problem}, hence if we show that $Q^k$ is an increasing function of $u^k$ (i.e., increasing in every coordinate $u_1^k, u_2^k, \ldots, u_n^k$), $\forall k$, then $(u_i^k)^*=\bar{u}_i^k$ is an optimum solution.

For a \emph{one-stage process} with initial state $x^{N-1}$, the $Q$-function is
\begin{equation}
\begin{split}
Q^{N-1}=\mu_1^{N-1}u_1^{N-1} +\ldots+\mu_n^{N-1}u_n^{N-1}.
\end{split}
\end{equation}
Since $\mu_i^k\geq 0$, $\forall i, k$, then $Q^{N-1}$ is an increasing function of $u_i^{N-1}$, $\forall i$, then $(u_i^{N-1})^*=\bar{u}_i^{N-1}$. 

For a \emph{two-stage process} with initial state $x^{N-2}$, the $Q$-function is
\begin{equation}\label{Q_N_2_GEN}
\begin{split}
Q^{N-2} &= (\mu_1^{N-2}u_1^{N-2} +\ldots+\mu_n^{N-2}u_n^{N-2}) + \\
&\;\;\;\;\,(\mu_1^{N-1}\bar{u}_1^{N-1} +\ldots+\mu_n^{N-1}\bar{u}_n^{N-1})\\
&= {\sum}_{i=1}^n(\mu_i^{N-2}u_i^{N-2}+\mu_i^{N-1}\bar{u}_i^{N-1}).
\end{split}
\end{equation}
From $x_i^{N-1}=x_i^{N-2}+T_s(y_i^{N-2}-u_i^{N-2})$, (\ref{case_1_junction}), and $y_i^k=R_{ri}u_r^k$, where $r$ is the only in-neighbor of cell $i$ (see (\ref{tot_inflow})),  
\begin{align}
\bar{u}_i^{N-1} &\!=
\min\Big{\{}\frac{v_i}{\ell_i}x_i^{N-1}, \Big{(}  \frac{w_{j}}{R_{ij}}(\gamma_{j}\!-\!\frac{1}{\ell_{j}}x_{j}^{N-1})\Big{)}_{j\in\mathcal{E}_i^+}, \nonumber\\
&\;\;\;\;\;\;\;C_i, \Big{(}   \frac{C_{j}}{R_{ij}}\Big{)}_{j\in\mathcal{E}_i^+}\!\Big{\}} \nonumber \\
&\!= \min\Big{\{}\frac{v_i}{\ell_i}x_i^{N-2}+\sigma_iR_{ri}u_r^{N-2}-\sigma_iu_i^{N-2}, \nonumber \\
&\;\;\;\;\;\;\Big{(}  \frac{w_{j}\gamma_{j}}{R_{ij}}\!-\!\frac{w_j}{R_{ij}\ell_{j}}x_{j}^{N-2} \!\!-\!\kappa_ju_i^{N-2} \!\!+\!\frac{\kappa_j}{R_{ij}}u_j^{N-2}\Big{)}_{j\in\mathcal{E}_i^+},\nonumber\\
&\;\;\;\;\;\;\;C_i, \Big{(}   \frac{C_{j}}{R_{ij}}\Big{)}_{j\in\mathcal{E}_i^+}\Big{\}},\label{aswth94ru61}
\end{align}
where $\sigma_i=(v_i/\ell_i)T_s\in[0,1]$, $\kappa_i=(w_i/\ell_i)T_s\in[0,1]$. By multiplying both sides of (\ref{aswth94ru61}) by $\mu_i^{N-1}$ and adding $\mu_i^{N-2}u_i^{N-2}$ to the both sides we obtain 
\begin{align}
&\mu_i^{N-2}u_i^{N-2}+\mu_i^{N-1}\bar{u}_i^{N-1}= \nonumber \\ &\min\!\Big{\{}\mu_i^{N-1}\frac{v_i}{\ell_i}x_i^{N-2}+\mu_i^{N-1}\sigma_iR_{ri}u_{r}^{N-2}+s_i^{N-2}u_i^{N-2}, \nonumber \\
&\;\;\;\;\;\;\;\;\;\Big{(}  \mu_i^{N-1}\frac{w_{j}\gamma_{j}}{R_{ij}}\!-\!\mu_i^{N-1}\frac{w_j}{R_{ij}\ell_{j}}x_{j}^{N-2} + \nonumber\\
&\;\;\;\;\;\;\;\;\;t_{ij}^{N-2}u_i^{N-2} \!+\!\mu_i^{N-1}\frac{\kappa_j}{R_{ij}}u_j^{N-2}\Big{)}_{j\in\mathcal{E}_i^+},\nonumber \\
&\;\;\;\;\;\;\;\;\;\mu_i^{N-1}C_i+\mu_i^{N-2}u_i^{N-2},\nonumber\\
&\;\;\;\;\;\;\;\;\;\Big{(}   \mu_i^{N-1}\frac{C_{j}}{R_{ij}}+\mu_i^{N-2}u_i^{N-2}\Big{)}_{j\in\mathcal{E}_i^+}\Big{\}}, \label{two_term_exp_GEN}
\end{align}
where $s_i^{N-2}=\mu_i^{N-2}-\mu_i^{N-1}\sigma_i$ and $t_{ij}^{N-2}=\mu_i^{N-2}-\mu_i^{N-1}\kappa_{j}$, $j\in\mathcal{E}_i^+$. Since $\mu_i^k\geq \mu_i^{k+1}\geq 0$ and $\sigma_i, \kappa_i\in[0,1]$, then $s_i^{N-2}, t_{ij}^{N-2}\geq 0$, $\forall i$. From (\ref{Q_N_2_GEN}) and (\ref{two_term_exp_GEN}), and that the coefficients of $u_i^{N-2}$ are non-negative $\forall i$, and using the fact that the minimum and the sum of increasing functions are also increasing, it follows that $Q^{N-2}$ is an increasing function of $u_i^{N-2}$, then $(u_i^{N-2})^*=\bar{u}_i^{N-2}$, $\forall i$, is an optimal control action.

For a $k$-\emph{stage process} with initial state $x^{N-k}$, assuming that $u_i^j=\bar{u}_i^j$, for $j=N-k+1, \ldots, N-2, N-1$, the $Q$-function is given by
\begin{equation}\label{Q_N_k_GEN}
\begin{split}
Q^{N-k} = 
{\sum}_{i=1}^n &(\mu_i^{N-k}u_i^{N-k}+\mu_i^{N-k+1}\bar{u}_i^{N-k+1}+\\
&\ldots+\mu_i^{N-2}\bar{u}_i^{N-2}+\mu_i^{N-1}\bar{u}_i^{N-1}).
\end{split}
\end{equation}
From (\ref{case_1_junction}) and that $x_i^{N-1}=x_i^{N-k}+T_s{\sum}_{l=N-k}^{N-2}(y_i^l-u_i^l)$, we have
\begin{align}
&\bar{u}_i^{N-1} \!=
\min\Big{\{}\frac{v_i}{\ell_i}x_i^{N-1}, \Big{(}  \frac{w_{j}}{R_{ij}}(\gamma_{j}\!-\!\frac{1}{\ell_{j}}x_{j}^{N-1})\Big{)}_{j\in\mathcal{E}_i^+}, \nonumber\\
&\;\;\;\;\;\;\;C_i, \Big{(}   \frac{C_{j}}{R_{ij}}\Big{)}_{j\in\mathcal{E}_i^+}\!\Big{\}} \nonumber \\
&\!= \min\Big{\{}\frac{v_i}{\ell_i}x_i^{N-k}+\sigma_iR_{ri}{\sum}_{l=N-k}^{N-2}u_r^{l}-\sigma_i{\sum}_{l=N-k}^{N-2}u_i^{l}, \nonumber \\
&\;\;\;\;\;\;\Big{(}  \frac{w_{j}\gamma_{j}}{R_{ij}}\!-\!\frac{w_j}{R_{ij}\ell_{j}}x_{j}^{N-k} \!\!-\!\kappa_j{\sum}_{l=N-k}^{N-2}u_i^{l} +\nonumber\\
&\;\;\;\;\;\;\frac{\kappa_j}{R_{ij}}{\sum}_{l=N-k}^{N-2}u_j^{l}\Big{)}_{j\in\mathcal{E}_i^+},C_i, \Big{(}   \frac{C_{j}}{R_{ij}}\Big{)}_{j\in\mathcal{E}_i^+}\Big{\}},\label{ygkuh645g58u08h98_GEN}
\end{align}
where $r$ is the only in-neighbor of cell $i$ and $u_i^j=\bar{u}_i^j$, for $j\geq N-k+1$. By multiplying both sides of (\ref{ygkuh645g58u08h98_GEN}) by $\mu_i^{N-1}$ and then adding ${\sum}_{l=N-k}^{N-2}\mu_i^lu_i^l$ to the both sides (wherein $u_i^l=\bar{u}_i^l$, for $l\geq N-k+1$), we obtain
\begin{align}
&{\sum}_{l=N-k}^{N-1}\mu_i^{l}u_i^{l}= \min\Big{\{}\mu_i^{N-1}\frac{v_i}{\ell_i}x_i^{N-k} + \label{ygkfgse11tyb8_GEN}\\
&\;\;\;\;\mu_i^{N-1}\sigma_iR_{ri}{\sum}_{l=N-k}^{N-2}\!u_{r}^l+{\sum}_{l=N-k}^{N-2}\!s_i^lu_i^{l},\nonumber \\
&\;\;\;\;\;\;\Big{(}  \mu_i^{N-1}\frac{w_{j}\gamma_{j}}{R_{ij}}\!-\!\mu_i^{N-1}\frac{w_j}{R_{ij}\ell_{j}}x_{j}^{N-k} \!+\!{\sum}_{l=N-k}^{N-2}t_{ij}^lu_i^{l} +\nonumber\\
&\;\;\;\;\;\;\mu_i^{N-1}\frac{\kappa_j}{R_{ij}}{\sum}_{l=N-k}^{N-2}u_j^{l}\Big{)}_{j\in\mathcal{E}_i^+}, \mu_i^{N-1}C_i + \nonumber\\
&\;\;\;\;\;\;{\sum}_{l=N-k}^{N-2}\mu_i^{l}u_i^{l}, \Big{(}   \mu_i^{N-1}\frac{C_{j}}{R_{ij}} \!+\! {\sum}_{l=N-k}^{N-2}\mu_i^{l}u_i^{l}\Big{)}_{j\in\mathcal{E}_i^+}\Big{\}}, \nonumber
\end{align}
where $s_i^l=\mu_i^l-\mu_i^{N-1}\sigma_i$ and $t_{ij}^l=\mu_i^l-\mu_i^{N-1}\kappa_{j}$, $j\in\mathcal{E}_i^+$. Since $\mu_i^k\geq \mu_i^{k+1}\geq 0$ and $\sigma_i, \kappa_i\in[0,1]$, then $s_i^l\geq s_i^{l+1}\geq 0$ and $t_{ij}^l\geq t_{ij}^{l+1}\geq 0$, $\forall i$ and any $l=N-k, \ldots, N-2$.

Due to the non-negativity of the coefficients of $u_i^j$ in the right-hand side of (\ref{ygkfgse11tyb8_GEN}),  ${\sum}_{l=N-k}^{N-1}\delta_i^{l}u_i^{l}$ is maximized if the terms ${\sum}_{l=N-k}^{N-2}\delta_{i-1}^{l}u_{i-1}^{l}$, ${\sum}_{l=N-k}^{N-2}\delta_i^{l}u_i^{l}$, and ${\sum}_{l=N-k}^{N-2}\delta_{i+1}^{l}u_{i+1}^{l}$ are maximized, $\forall i$, and so on, where $\delta_i^l$ be a generic symbol for a sequence of parameters satisfying $\delta_i^l\geq \delta_i^{l+1}\geq 0$, $\forall i, l$.  The above recursion implies that $(u_i^{N-k})^*=\bar{u}_i^{N-k}$, $\forall i$, is an optimal solution.  

Therefore, the optimum control is independent of the control horizon $N$ and external inflow rates $\lambda_i^k$, and is obtained at no computational cost by setting each $u_i^k$ equal to its known upper limit $\bar{u}_i^k$,  $\forall i, k$ (given that $x^k$ is known at time $k$). From the expression for $\bar{u}_i^k$, it follows that the true optimal outflow rate $u_i^k$ is found by measuring only $x_i^k$ and $\left(x_{j}^k\right)_{j\in\mathcal{E}_i^+}$.  $\hfill\blacksquare$

\emph{Proof of Theorem~\ref{thm_GEN_DENTTRAFF}}:
Since the decentralized controller is obtained by solving a centralized problem for a sub-network consisting of cell $i$ and $j\in\mathcal{D}_i$, then the proofs follows from Theorem~\ref{thm1} and Corollary~\ref{cor01}. The feasibility of $(\hat{u}_i^k)^*$ in (\ref{suboptimal_i}), $\forall i, k$, follows from the proof of Theorem~\ref{thm1} and that from (\ref{case_1_junction}) and (\ref{case_2_junction}), at each time $k$, the feasibility of $u_i^k$ depends on the values of $x_i^k$ and $x_j^k$, $\forall j\in\mathcal{D}_i$ and is independent of the state of non-local variables. Hence, the feasibility of $u_i^k$ is ensured in the $i$-th local optimization problem.  $\hfill\blacksquare$


\bibliography{myRefs}

\begin{thebibliography}{10}
\providecommand{\url}[1]{#1}
\csname url@samestyle\endcsname
\providecommand{\newblock}{\relax}
\providecommand{\bibinfo}[2]{#2}
\providecommand{\BIBentrySTDinterwordspacing}{\spaceskip=0pt\relax}
\providecommand{\BIBentryALTinterwordstretchfactor}{4}
\providecommand{\BIBentryALTinterwordspacing}{\spaceskip=\fontdimen2\font plus
\BIBentryALTinterwordstretchfactor\fontdimen3\font minus
  \fontdimen4\font\relax}
\providecommand{\BIBforeignlanguage}[2]{{%
\expandafter\ifx\csname l@#1\endcsname\relax
\typeout{** WARNING: IEEEtran.bst: No hyphenation pattern has been}%
\typeout{** loaded for the language `#1'. Using the pattern for}%
\typeout{** the default language instead.}%
\else
\language=\csname l@#1\endcsname
\fi
#2}}
\providecommand{\BIBdecl}{\relax}
\BIBdecl

\bibitem{Munson12}
B.~R. Munson, A.~P. Rothmayer, T.~H. Okiishi, and W.~W. Huebsch,
  \emph{Fundamentals of Fluid Mechanics}.\hskip 1em plus 0.5em minus
  0.4em\relax John Wiley \& Sons, Inc., 2012.

\bibitem{Como16}
G.~Como, E.~Lovisari, and K.~Savla, ``Convexity and robustness of dynamic
  traffic assignment and freeway network control,'' \emph{Transportation
  Research Part B: Methodological}, vol.~91, pp. 446--465, 2016.

\bibitem{Han2017}
Y.~Han, A.~Hegyi, Y.~Yuan, S.~Hoogendoorn, M.~Papageorgiou, and C.~Roncoli,
  ``Resolving freeway jam waves by discrete first-order model-based predictive
  control of variable speed limits,'' \emph{Transportation Research Part C:
  Emerging Technologies}, vol.~77, pp. 405--420, 2017.

\bibitem{Muralidharan2015}
A.~Muralidharan and R.~Horowitz, ``Computationally efficient model predictive
  control of freeway networks,'' \emph{Transportation Research Part C: Emerging
  Technologies}, vol.~58, pp. 532--553, 2015.

\bibitem{Daganzo94}
C.~Daganzo, ``The cell transmission model: {A} dynamic representation of
  highway traffic consistent with the hydrodynamic theory,''
  \emph{Transportation Research Part B}, vol.~28, no.~4, pp. 269--287, 1994.

\bibitem{Adacher2018}
L.~Adacher and M.~Tiriolo, ``A macroscopic model with the advantages of
  microscopic model: {A} review of cell transmission model's extensions for
  urban traffic networks,'' \emph{Simulation Modelling Practice and Theory},
  vol.~86, pp. 102--119, 2018.

\bibitem{Wong1968}
P.~Wong and R.~Larson, ``Optimization of natural-gas pipeline systems via
  dynamic programming,'' \emph{IEEE Transactions on Automatic Control},
  vol.~13, no.~5, pp. 475--481, 1968.

\bibitem{Misra15}
S.~Misra, M.~W. Fisher, S.~Backhaus, R.~Bent, M.~Chertkov, and F.~Pan,
  ``Optimal compression in natural gas networks: A geometric programming
  approach,'' \emph{IEEE Transactions on Control of Network Systems}, vol.~2,
  no.~1, pp. 47--56, 2015.

\bibitem{Martin2006}
A.~Martin, M.~M{\"o}ller, and S.~Moritz, ``Mixed integer models for the
  stationary case of gas network optimization,'' \emph{Mathematical
  Programming}, vol. 105, no.~2, pp. 563--582, 2006.

\bibitem{Koch2015}
T.~Koch, B.~Hiller, M.~Pfetsch, and L.~Schewe, \emph{Evaluating Gas Network
  Capacities}.\hskip 1em plus 0.5em minus 0.4em\relax Philadelphia, PA: Society
  for Industrial and Applied Mathematics, 2015.

\bibitem{Hegyi2005}
A.~Hegyi, B.~D. Schutter, and H.~Hellendoorn, ``Model predictive control for
  optimal coordination of ramp metering and variable speed limits,''
  \emph{Transportation Research Part C: Emerging Technologies}, vol.~13, no.~3,
  pp. 185--209, 2005.

\bibitem{Papamichail2010}
I.~Papamichail, A.~Kotsialos, I.~Margonis, and M.~Papageorgiou, ``Coordinated
  ramp metering for freeway networks-- {A} model-predictive hierarchical
  control approach,'' \emph{Transportation Research Part C: Emerging
  Technologies}, vol.~18, no.~3, pp. 311--331, 2010.

\bibitem{Hadiuzzaman2013}
M.~Hadiuzzaman and T.~Z. Qiu, ``Cell transmission model based variable speed
  limit control for freeways,'' \emph{Canadian Journal of Civil Engineering},
  vol.~40, no.~1, pp. 46--56, 2013.

\bibitem{Tsitsiklis_Athans_1985}
J.~Tsitsiklis and M.~Athans, ``On the complexity of decentralized decision
  making and detection problems,'' \emph{IEEE Transactions on Automatic
  Control}, vol.~30, no.~5, pp. 440--446, 1985.

\bibitem{Cogill2006}
R.~Cogill, M.~Rotkowitz, B.~V. Roy, and S.~Lall, ``An approximate dynamic
  programming approach to decentralized control of stochastic systems,'' in
  \emph{Control of Uncertain Systems: Modelling, Approximation, and
  Design}.\hskip 1em plus 0.5em minus 0.4em\relax Springer Berlin Heidelberg,
  2006, pp. 243--256.

\bibitem{Lakshmanan2006}
H.~Lakshmanan and D.~P. de~Farias, ``Decentralized approximate dynamic
  programming for dynamic networks of agents,'' in \emph{2006 American Control
  Conference (ACC)}, June 2006, pp. 1648--1653.

\bibitem{Borrelli03}
F.~Borrelli, \emph{Constrained Optimal Control of Linear and Hybrid
  Systems}.\hskip 1em plus 0.5em minus 0.4em\relax Springer, 2003.

\bibitem{MPT3}
\BIBentryALTinterwordspacing
M.~Herceg, M.~Kvasnica, C.~N. Jones, and M.~Morari, ``{Multi-Parametric Toolbox
  3.0},'' in \emph{Proc.~of the European Control Conference}, Z\"urich,
  Switzerland, July 17--19 2013, pp. 502--510. [Online]. Available:
  \url{http://control.ee.ethz.ch/~mpt}
\BIBentrySTDinterwordspacing

\bibitem{YALMIP_site}
\BIBentryALTinterwordspacing
J.~Lofberg. {YALMIP}: {A} toolbox for modeling and optimization in {MATLAB}.
  [Online]. Available: \url{https://yalmip.github.io/download/}
\BIBentrySTDinterwordspacing

\bibitem{Krumm10}
J.~Krumm, ``Where will they turn: predicting turn proportions at
  intersections,'' \emph{Personal and Ubiquitous Computing}, vol.~14, no.~7,
  pp. 591--599, 2010.

\bibitem{Baotic2003}
M.~Baotic, F.~J. Christophersen, and M.~Morari, ``A new algorithm for
  constrained finite time optimal control of hybrid systems with a linear
  performance index,'' in \emph{2003 European Control Conference (ECC)}, 2003,
  pp. 3323--3328.

\bibitem{Bertsimas2006}
D.~Bertsimas and A.~Thiele, ``A robust optimization approach to inventory
  theory,'' \emph{Operations Research}, vol.~54, no.~1, pp. 150--168, 2006.

\bibitem{Gabrel2010}
V.~Gabrel, C.~Murat, and N.~Remli, ``Linear programming with interval right
  hand sides,'' \emph{International Transactions in Operational Research},
  vol.~17, no.~3, pp. 397--408, 2010.

\bibitem{Chinneck2000}
J.~W. Chinneck and K.~Ramadan, ``Linear programming with interval
  coefficients,'' \emph{The Journal of the Operational Research Society},
  vol.~51, no.~2, pp. 209--220, 2000.

\bibitem{Wong1968b}
P.~Wong and R.~Larson, ``Optimization of tree-structured natural-gas
  transmission networks,'' \emph{Journal of Mathematical Analysis and
  Applications}, vol.~24, no.~3, pp. 613--626, 1968.

\bibitem{Wolf2000}
D.~D. Wolf and Y.~Smeers, ``The gas transmission problem solved by an extension
  of the simplex algorithm,'' \emph{Management Science}, vol.~46, no.~11, pp.
  1454--1465, 2000.

\bibitem{Zlotnik2015}
A.~{Zlotnik}, M.~{Chertkov}, and S.~{Backhaus}, ``Optimal control of transient
  flow in natural gas networks,'' in \emph{2015 54th IEEE Conference on
  Decision and Control (CDC)}, Dec 2015, pp. 4563--4570.

\bibitem{Stefanov2001}
S.~M. Stefanov, \emph{Separable Programming: Theory and Methods}.\hskip 1em
  plus 0.5em minus 0.4em\relax Springer, Boston, MA, 2001.

\bibitem{Jensen2003}
P.~A. Jensen and J.~F. Bard, \emph{Operations Research Models and
  Methods}.\hskip 1em plus 0.5em minus 0.4em\relax John Wiley \& Sons, Inc.,
  2003.

\bibitem{Hager2016}
W.~W. Hager, D.~T. Phan, and J.~Zhu, ``Projection algorithms for nonconvex
  minimization with application to sparse principal component analysis,''
  \emph{Journal of Global Optimization}, vol.~65, no.~4, pp. 657--676, 2016.

\bibitem{Rockafellar70}
R.~T. Rockafellar, \emph{Convex Analysis}.\hskip 1em plus 0.5em minus
  0.4em\relax Princeton University Press, 1970.

\bibitem{Horst2000}
R.~Horst, P.~M. Pardalos, and N.~V. Thoai, \emph{Introduction to Global
  Optimization}.\hskip 1em plus 0.5em minus 0.4em\relax Springer, US, 2000.

\bibitem{Borrelli_Thesis_2002}
F.~Borrelli, ``Discrete time constrained optimal control,'' Ph.D. dissertation,
  Swiss Federal Institute of Technology (ETH) Zurich, 2002.

\bibitem{Frank_Lewis12}
F.~L. Lewis, D.~L. Vrabie, and V.~L. Syrmos, \emph{Optimal Control}.\hskip 1em
  plus 0.5em minus 0.4em\relax John Wiley \& Sons, Inc., 2012.

\end{thebibliography}

\end{document}